\documentclass[11pt]{article}
\newif\ifarxiv
\arxivtrue
\ifarxiv\usepackage[preprint]{acl}\else\usepackage[review]{acl}\fi
\usepackage{times,latexsym}
\usepackage[utf8]{inputenc}
\usepackage[T1]{fontenc}
\usepackage{amsmath,amssymb,booktabs,array,graphicx,microtype,xcolor,xspace}
\usepackage{xurl}
\ifarxiv
\hypersetup{pdftitle={Can Agentic Code Review Catch Vulnerabilities Before They Are Merged?},pdfauthor={Yihan Yang, Chenyuan Yang, Zijie Zhao, Anh Nguyen Hoang, Ziqi Zhang, Weidong Wang, Zhiheng Lyu, Lingming Zhang}}
\else
\hypersetup{pdftitle={Can Agentic Code Review Catch Vulnerabilities Before They Are Merged?},pdfauthor={Anonymous authors}}
\fi
\usepackage{tikz,pgfplots}
\usetikzlibrary{arrows.meta,positioning}
\usepackage{fontawesome5}
\ifdefined\pdfmapfile\pdfmapfile{=fontawesome5.map}\fi
\definecolor{agentstroke}{RGB}{0,114,178}
\definecolor{evalfill}{RGB}{252,242,220}
\definecolor{evalstroke}{RGB}{176,112,0}
\definecolor{diffadd}{RGB}{230,255,237}
\definecolor{diffdel}{RGB}{255,235,233}
\definecolor{okgreen}{RGB}{0,158,115}
\definecolor{badred}{RGB}{213,94,0}
\definecolor{warnamber}{RGB}{230,159,0}
\pgfplotsset{compat=1.18}
\usepackage{placeins}
\usepackage{enumitem}
\usepackage{pifont,colortbl,makecell}
\DeclareRobustCommand{\cmark}{\textcolor{green!45!black}{\ding{51}}}
\DeclareRobustCommand{\xmark}{\textcolor{black!35}{\ding{55}}}
\DeclareRobustCommand{\pmark}{\tikz[baseline=-0.6ex]{\draw[black!60,line width=0.5pt] (0,0) circle (0.9ex); \fill[black!60] (0,0.9ex) arc (90:270:0.9ex) -- cycle;}}
\newcommand{\tcite}[1]{{\scriptsize\citep{#1}}}
\newcommand{\rowgroup}[1]{\multicolumn{9}{@{}l}{\scriptsize\textsc{\textcolor{black!60}{#1}}}\\[-1pt]}
\newcommand{\rowgroupn}[2]{\multicolumn{#1}{@{}l}{\scriptsize\textsc{\textcolor{black!60}{#2}}}\\[-1pt]}
\definecolor{oursrow}{RGB}{232,240,252}
\newcommand{\claudecodeicon}{\raisebox{-0.15em}{\includegraphics[height=1.0em]{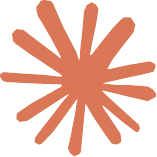}}}
\newcommand{\codexicon}{\raisebox{-0.15em}{\includegraphics[height=1.0em]{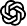}}}
\newcommand{\dropcell}[1]{\textcolor{okgreen}{\scalebox{0.75}{\faArrowDown}}\,#1\%}
\newcommand{\risecell}[1]{\textcolor{black!70}{\scalebox{0.75}{\faArrowUp}}\,#1\%}
\definecolor{heatblue}{HTML}{2A78D6}
\newcommand{\heat}[2]{\cellcolor{heatblue!#1!white}#2}
\newcommand{\bench}{SecPRBench\xspace}
\newcommand{\code}[1]{\texttt{#1}}
\providecommand{\cyy}[1]{}
\providecommand{\yyh}[1]{}

\providecommand{\TODO}[1]{}

\newcommand{\finalVerifiedCases}{265}
\newcommand{\finalVerifiedInputs}{530}

\newcommand{\finalMechanicalVerifiedCases}{217}
\newcommand{\finalHumanVerifiedCases}{48}

\newcommand{\numBuggy}{265}

\newcommand{\numFixed}{265}
\newcommand{\numProjects}{nine}
\newcommand{\medianFiles}{4}
\newcommand{\medianLines}{173}
\newcommand{\judgeModel}{Claude Opus~5.5 (high)}
\newcommand{\recallLow}{32.5\%}
\newcommand{\recallHigh}{78.1\%}

\newcommand{\numJudgeAudit}{50}

\newcommand{\numJudgeStabilityReports}{2,143}
\newcommand{\judgeStabilityAgreement}{96.9\%}

\newcommand{\sNeverFound}{23}



\title{Can Agentic Code Review Catch Vulnerabilities Before They Are Merged?}
\ifarxiv
\author{
 \textbf{Yihan Yang\thanks{Equal contribution.}},
 \textbf{Chenyuan Yang\footnotemark[1]},
 \textbf{Zijie Zhao},
 \textbf{Anh Nguyen Hoang\thanks{Quantum AI \& Cyber Security Institute, FPT Corporation; work done during an internship at UIUC.}},
\\
 \textbf{Ziqi Zhang},
 \textbf{Weidong Wang},
 \textbf{Zhiheng Lyu},
 \textbf{Lingming Zhang}
\\
 \raisebox{-0.22em}{\includegraphics[height=1.1em]{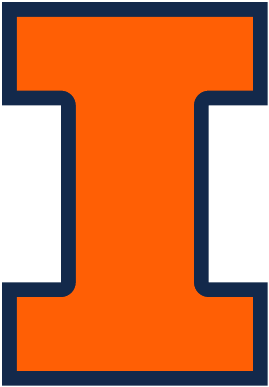}}\hspace{0.3em}University of Illinois Urbana-Champaign
\\
 \texttt{\{yihany5, cy54, lingming\}@illinois.edu}
}
\else
\author{Anonymous authors}
\fi

\begin{document}
\maketitle
\begin{abstract}
LLM agents have been increasingly adopted for security-focused code review, but existing benchmarks mostly derive the reviewed changes from their later fixes instead of the real-world bug-introducing commits. In this paper, we present \textbf{\bench}, the first comprehensive benchmark for evaluating agentic security review on validated real-world code changes. More specifically, we have built an automated agentic pipeline to autonomously harvest, build, and validate massive real-world bug-introducing commits for popular software systems. To ensure precise evaluation, we also adopt LLM-as-a-judge to credit the findings matching ground-truth vulnerabilities.
%a benchmark that gives an agent the real change that introduced a vulnerability, with its repository but no description of the flaw.
To date, we have collected \numBuggy{} tasks from \numProjects{} critical systems, including Firefox, V8, and OpenSSL with millions of lines of code.
%An automated pipeline builds and verifies these pairs, mostly by proof-of-concept replay, and supports refreshes from new disclosures.
%A diff already shows a reviewer where to look, so an LLM judge credits only findings that explain the vulnerability's root cause in the reviewed code.
Across eight frontier models, recall ranges from \recallLow{} to \recallHigh{}, with Grok 4.6 and GPT-6 sol ranking at the top. Our study also shows that most misses are not simply oversights: in 68\% of them, the agent reports on a file that holds the flaw but argues for a different mechanism. Security-focused harnesses also behave differently: OpenAI's Codex Security lowers recall by 8.3 points, mostly by abstaining, while Anthropic's Defending Code harness raises it by 7.7 points, at a higher cost.

\ifarxiv
\par\smallskip\centerline{\small\faGithub~\href{https://github.com/secprbench/secprbench}{\texttt{github.com/secprbench/secprbench}}}
\centerline{\small\faGlobe~\href{https://secprbench.github.io}{\texttt{secprbench.github.io}}}
\fi
\end{abstract}
\section{Introduction}
\label{sec:introduction}

Anthropic, OpenAI, and GitHub now offer agents that review code changes, such as pull requests (PRs), for vulnerabilities before merge~\citep{anthropicPRSecurityReview,openai2025codexReview,github2026securityReview}.
A vulnerability missed at review can persist for years~\citep{yang2025knighter}.
Consider a Firefox sandbox escape in which a compromised child process could make the privileged parent return an unintentionally powerful handle over inter-process communication (IPC)~\citep[\mbox{CVE-2025-2857};][]{mozilla2025sandboxescape}.
Catching it at review requires connecting the IPC code to Windows handle semantics and the trust boundary between processes.
We ask: \emph{Can today's agents catch vulnerabilities before merge?}
% Figure 1: one SecPRBench case end to end (buggy task + fixed companion), the
% agent/evaluator information boundary, and the judge's two-criterion verdicts.
% Dataset-level counts are deliberately absent (they live in Sections 3.2-3.5);
% the only numbers are the example's own diffstats.
% Preamble requirements (main.tex): \usepackage{fontawesome5}; tikz library arrows.meta;
% colours oursrow, agentstroke, evalfill, evalstroke (RGB 176,112,0: grayscale-distinct from the
% blue border and away from the Section 5 figure's companion orange), diffadd, diffdel, okgreen, badred.
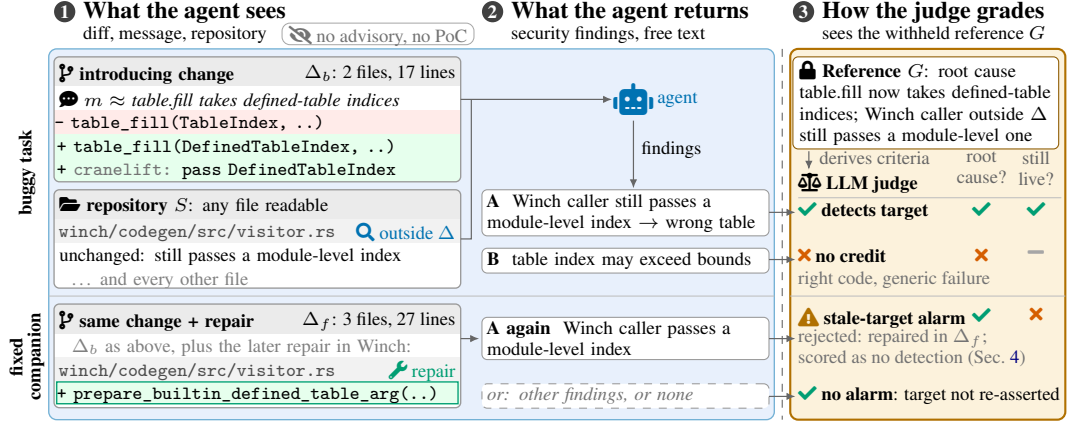
\begin{figure*}[t]
\centering
\begin{tikzpicture}[x=1pt,y=-1pt,font=\scriptsize,
  every node/.style={inner sep=0pt,outer sep=0pt},
  tw/.style={anchor=north west,align=left},
  row/.style={anchor=north west,inner xsep=2pt,inner ysep=0pt,minimum height=9pt,
              text width=149pt,align=left,text depth=0pt},
  code/.style={row,font=\scriptsize\fontfamily{lmtt}\selectfont},
  tag/.style={anchor=north east,inner xsep=2pt,minimum height=9pt,text depth=0pt},
  chip/.style={anchor=north west,draw=black!40,fill=white,rounded corners=2pt,line width=0.4pt,
               inner xsep=2pt,inner ysep=1.5pt,text width=104pt,align=left},
  vlab/.style={anchor=north west,align=left},
  glyph/.style={anchor=north,inner sep=0pt},
  hdr/.style={anchor=north west,align=left,font=\footnotesize\bfseries},
  num/.style={circle,fill=black!75,text=white,minimum size=8pt,font=\scriptsize\bfseries,inner sep=0pt},
  arr/.style={-{Latex[length=1.2mm,width=1mm]},line width=0.45pt,black!60},
  pill/.style={draw=black!35,fill=white,rounded corners=3pt,line width=0.35pt,
               inner xsep=2.5pt,inner ysep=1pt,text depth=0pt,text=black!55},
  ]
% ================================================================ regions (drawn first)
\draw[fill=oursrow,draw=agentstroke!55,line width=0.5pt,rounded corners=3pt] (15,18) rectangle (288,157.5);
\draw[fill=evalfill,draw=evalstroke,line width=0.8pt,rounded corners=3pt] (294,18) rectangle (397.5,157.5);
\draw[black!50,dashed,line width=0.5pt] (291,18) -- (291,157.5);
\draw[black!25,line width=0.4pt] (15,111) -- (288,111);
\draw[black!25,line width=0.4pt] (294,111) -- (397.5,111);
% ================================================================ headers
\node[num] at (21,4) {1};
\node[hdr] at (28,0) {What the agent sees};
\node[tw] at (28,9.5) {diff, message, repository};
\node[pill,anchor=north east] at (175,8.3) {\faEyeSlash\ no advisory, no PoC};
\node[num] at (182,4) {2};
\node[hdr] at (189,0) {What the agent returns};
\node[tw] at (189,9.5) {security findings, free text};
\node[num] at (299,4) {3};
\node[hdr] at (306,0) {How the judge grades};
\node[tw] at (306,9.5) {sees the withheld reference $G$};
% ================================================================ row labels
\node[rotate=90,anchor=center,font=\scriptsize\bfseries,align=center] at (6.8,64.5) {buggy task};
\node[rotate=90,anchor=center,font=\scriptsize\bfseries\linespread{0.85}\selectfont,align=center] at (6.8,133.75) {fixed\\companion};
% ================================================================ buggy row: PR card (x 17..170, y 21..68)
\begin{scope}
  \clip[rounded corners=2pt] (17,21) rectangle (170,68);
  \fill[white] (17,21) rectangle (170,68);
  \node[row,minimum height=11pt,fill=black!8] at (17,21) {\faCodeBranch\ \textbf{introducing change}};
  \node[tag,minimum height=11pt] at (170,21) {$\Delta_b$: 2 files, 17 lines};
  \node[row] at (17,32) {\faCommentDots\ $m\approx$ \emph{table.fill takes defined-table indices}};
  \node[code,fill=diffdel] at (17,41) {- table\_fill(TableIndex, ..)};
  \node[code,fill=diffadd] at (17,50) {+ table\_fill(DefinedTableIndex, ..)};
  \node[code,fill=diffadd] at (17,59) {+ \textcolor{black!55}{cranelift:} pass DefinedTableIndex};
\end{scope}
\draw[black!40,line width=0.5pt,rounded corners=2pt] (17,21) rectangle (170,68);
% ================================================================ buggy row: repository card (x 17..170, y 71..108)
\begin{scope}
  \clip[rounded corners=2pt] (17,71) rectangle (170,108);
  \fill[white] (17,71) rectangle (170,108);
  \node[row,minimum height=11pt,fill=black!8] at (17,71) {\faFolderOpen\ \textbf{repository $S$}: any file readable};
  \node[code,fill=black!5,text=black!70] at (17,82) {winch/codegen/src/visitor.rs};
  \node[tag,text=agentstroke] at (170,82) {\faSearch\ outside $\Delta$};
  \node[row] at (17,91) {unchanged: still passes a module-level index};
  \node[row,text=black!50] at (17,100) {\ \ \ldots\ and every other file};
\end{scope}
\draw[black!40,line width=0.5pt,rounded corners=2pt] (17,71) rectangle (170,108);
% ================================================================ buggy row: agent and findings
% the repository card also feeds the agent: merge connector in the card/chip gutter
\draw[black!60,line width=0.45pt] (170,89.5) -- (174,89.5) -- (174,37);
\draw[arr] (170,37) -- (227,37);
\node[font=\large,text=agentstroke,anchor=center] (rb) at (235,37) {\faRobot};
\node[anchor=west,text=agentstroke] at (244,37) {agent};
\draw[arr] (235,45) -- (235,70.5);
\node[anchor=west] at (238,56) {findings};
\node[chip] (f1) at (178,71) {\textbf{A}\ \ Winch caller still passes a\\ module-level index $\to$ wrong table};
\node[chip] (f2) at (178,93) {\textbf{B}\ \ table index may exceed bounds};
\draw[arr] (286,79.5) -- (296,79.5);
\draw[arr] (286,97.25) -- (296,97.25);
% ================================================================ buggy row: evaluator column
\node[anchor=north west,draw=evalstroke!80!black,fill=white,rounded corners=2pt,line width=0.4pt,inner xsep=2.5pt,inner ysep=1.5pt,text width=95pt,align=left] (g) at (295,21)
  {\faLock\ \textbf{Reference $G$}: root cause\\ table.fill now takes defined-table indices; Winch caller outside $\Delta$ still passes a module-level one};
\draw[arr] (301,56.5) -- (301,63.5);
\node[tw,text=black!60] at (305,56.5) {derives criteria};
\node[glyph,align=center,text=black!60] at (366,56.5) {root\\cause?};
\node[glyph,align=center,text=black!60] at (386.5,56.5) {still\\live?};
\node[tw] at (297,64.5) {\faBalanceScale\ \textbf{LLM judge}};
\draw[black!25,line width=0.4pt] (297,74) -- (396,74);
\node[vlab] at (297,75.5) {\textcolor{okgreen}{\faCheck}\ \textbf{detects target}};
\node[glyph,text=okgreen] at (366,75.5) {\faCheck};
\node[glyph,text=okgreen] at (386.5,75.5) {\faCheck};
\node[vlab] at (297,93.25) {\textcolor{badred}{\faTimes}\ \textbf{no credit}};
\node[glyph,text=badred] at (366,93.25) {\faTimes};
\node[glyph,text=black!35] at (386.5,93.75) {\faMinus};
\node[tw,text=black!55] at (297,101.75) {right code, generic failure};
% ================================================================ fixed row (y 114..153)
\begin{scope}
  \clip[rounded corners=2pt] (17,114) rectangle (170,153);
  \fill[white] (17,114) rectangle (170,153);
  \node[row,minimum height=11pt,fill=black!8] at (17,114) {\faCodeBranch\ \textbf{same change + repair}};
  \node[tag,minimum height=11pt] at (170,114) {$\Delta_f$: 3 files, 27 lines};
  \node[row,text=black!50] at (17,125) {\ \ $\Delta_b$ as above, plus the later repair in Winch:};
  \node[code,fill=black!5,text=black!70] at (17,134) {winch/codegen/src/visitor.rs};
  \node[tag,text=okgreen] at (170,134) {\faWrench\ repair};
  \node[code,fill=diffadd] at (17,143) {+ prepare\_builtin\_defined\_table\_arg(..)};
  \draw[okgreen,line width=0.5pt] (18,143.3) rectangle (169,151.5);
\end{scope}
\draw[black!40,line width=0.5pt,rounded corners=2pt] (17,114) rectangle (170,153);
\node[chip] (f3) at (178,119) {\textbf{A again}\ \ Winch caller passes a module-level index};
\node[chip,dashed,text=black!55] (f4) at (178,144) {\emph{or: other findings, or none}};
\draw[arr] (170,127) -- (177,127);
\draw[arr] (286,127) -- (296,127);
\draw[arr] (286,148.75) -- (296,148.75);
\node[vlab] at (297,115.5) {\textcolor{evalstroke}{\faExclamationTriangle}\ \textbf{stale-target alarm}};
\node[glyph,text=okgreen] at (366,115.5) {\faCheck};
\node[glyph,text=badred] at (386.5,115.5) {\faTimes};
\node[tw,text=black!55] at (297,123.5) {rejected: repaired in $\Delta_f$;};
\node[tw,text=black!55] at (297,131.5) {scored as no detection (Sec.~\ref{sec:evaluation})};
\node[vlab] at (297,144.75) {\textcolor{okgreen}{\faCheck}\ \textbf{no alarm}: target not re-asserted};
\end{tikzpicture}
\caption{\textbf{One \bench case} (Wasmtime CVE-2026-34946; diff and message paraphrased, findings \mbox{illustrative}). The flaw depends on unchanged code outside the diff; the LLM judge credits a finding only if it explains the target's root cause and that cause is still live in the reviewed code.}
\label{fig:overview}
\end{figure*}

A diff shows a reviewer where to start looking but not what is wrong, so a security review is only as useful as its diagnosis.
Unfortunately, existing benchmarks do not measure diagnosis on real changes (Table~\ref{tab:related}).
Earlier detection datasets score a binary label on single functions~\citep{ding2025primevul,yildiz2025jitvul}, which models often get right for the wrong reasons~\citep{ullah2024llms}.
Reproduction and exploitation benchmarks start from a known vulnerability and score an executable proof-of-concept (PoC) or exploit~\citep{wang2026cybergym,lee2026secbenchpro,wang2026exploitgym,lee2026exploitbench}.
Code-review benchmarks also do not target security vulnerabilities~\citep{zeng2026swrbench,zhang2026crab}.
Security PR-review benchmarks come closest, and MalPR-Bench scores whether the reviewer identifies the target~\citep{chen2026verdict}.
However, most of their vulnerable PRs revert or partially apply a later fix, which centers each diff on the repaired code, instead of using the real vulnerability-introducing commit~\citep{melo2026sevra,chen2026verdict}.
%Measuring diagnosis needs three things: the change that introduced the flaw, shown without hints; the repaired change as a control; and a score for the explanation.

We introduce \bench, a benchmark for \emph{pre-merge security review} that aims to evaluate modern agentic code review on real-world bug-introducing commits.
Each task gives an agent the diff of a real historical change that introduced a vulnerability, its commit message, and the repository, but no description of the flaw; the agent reports \emph{findings}.
A finding earns credit only if it explains the root cause of the task's \emph{target} vulnerability in the reviewed code; in Figure~\ref{fig:overview}, a bounds warning on the right call earns none.
\bench contains \numBuggy{} buggy tasks from \numProjects{} widely deployed C, C++, and Rust projects, including Firefox, V8, and OpenSSL with millions of lines of code.
Each buggy task has a \emph{fixed companion} that adds the later repair, on which the target should no longer be reported.

Building such pairs is challenging as advisories typically name the fix, not the introducing change, and automatic attribution can be unreliable~\citep{lu2026vicbench}.
We have built an automated agentic pipeline to trace each bug's introducing change and compose it with the later fix. Moreover, our pipeline also validates the pairs automatically, mostly by PoC replay on the parent, introducing, and fixed versions.
Most cases pass mechanical validation checks based on PoC execution; the other task instances require human adjudication.
For task evaluation, since no oracle checks free text, a large language model (LLM) judge, \judgeModel{}, derives target-specific criteria from a reference built from public security records and checks each finding against them and the reviewed code.

\looseness=-1
Across eight frontier models and two security-focused harnesses, three findings stand out.
\textbf{Strong agents tie on easy targets but diverge on contested ones.}
Grok~4.6, GPT-6 sol, GLM~5.3, and Opus~5.5 each find about three in four targets and are statistically tied overall.
On easy targets, which nearly all other agents find, every agent but MiniMax~M3 finds at least nine in ten.
On contested targets, which few other agents find, the top four still find about half, while the other four find at most one in five.
Even combined, the eight models leave \sNeverFound{} targets unfound.
\textbf{Most misses are not oversights.}
In 68\% of misses, a finding lands in a file that holds the flaw, often in the very function, but argues a mechanism the judge rejects.
Scoring by location or by a binary vulnerable/safe verdict would count these as detections; MalPR-Bench similarly finds that reviewers often block a PR without identifying its flaw~\citep{chen2026verdict}.
\textbf{Security-focused harnesses move recall in both directions.}
With each backend model fixed, OpenAI's Codex Security plugin, run as a static diff scan~\citep{openai2026codexSecurityDocs}, lowers recall by 8.3 points relative to the same model in the Codex CLI, mostly by abstaining; Anthropic's Defending Code reference harness~\citep{anthropicDefendingCodeHarness} raises it by 7.7 points on the tasks it completed, at higher cost and report volume.
On repaired changes, the judge never rules that the removed target is still present (\S~\ref{sec:fixed-results}).

% Symbol macros (\cmark, \xmark, \pmark, \tcite, \rowgroup) are defined in main.tex.
\begin{table*}[t]
\centering

\footnotesize
\setlength{\tabcolsep}{4.2pt}
\renewcommand{\arraystretch}{1.0}
\begin{tabular}{@{}l r ccc cc l c@{}}
\toprule
& & \multicolumn{3}{c}{\textbf{Task input}} & \multicolumn{2}{c}{\textbf{Validation}} & \multicolumn{2}{c}{\textbf{Scoring}} \\
\cmidrule(lr){3-5}\cmidrule(lr){6-7}\cmidrule(l){8-9}
\textbf{Benchmark} & \textbf{Tasks} & \makecell{Real\\change} & \makecell{No\\hint} & \makecell{Full\\repo} & \makecell{Paired\\control} & \makecell{PoC-\\validated} & \textbf{Scored by} & \makecell{Root\\cause} \\
\midrule
\rowgroup{Vulnerability detection: label}
PrimeVul \tcite{ding2025primevul}          & 6{,}968 & \xmark & \cmark & \xmark & \cmark & \xmark & Label          & \xmark \\
JitVul \tcite{yildiz2025jitvul}            & 1{,}758 & \pmark & \cmark & \pmark & \cmark & \xmark & Label          & \xmark \\
\addlinespace[3pt]
\rowgroup{Reproduction and exploitation: PoC, exploit, or patch}
SEC-bench \tcite{lee2025secbench}          & 200     & \xmark & \xmark & \cmark & \xmark & \cmark & Exec.          & \xmark \\
CyberGym \tcite{wang2026cybergym}          & 1{,}507 & \xmark & \pmark & \cmark & \xmark & \cmark & Exec.          & \xmark \\
CyberGym-E2E \tcite{shi2026cybergyme2e}    & 920     & \xmark & \cmark & \cmark & \xmark & \cmark & Exec.          & \xmark \\
SEC-bench Pro \tcite{lee2026secbenchpro}   & 344     & \xmark & \pmark & \cmark & \xmark & \cmark & Exec.\,+\,LLM  & \xmark \\
ExploitGym \tcite{wang2026exploitgym}      & 898     & \xmark & \xmark & \cmark & \xmark & \cmark & Exec.\,+\,LLM  & \xmark \\
ExploitBench \tcite{lee2026exploitbench}   & 41      & \xmark & \xmark & \cmark & \xmark & \cmark & Exec.          & \xmark \\
\addlinespace[3pt]
\rowgroup{Code review: comments}
c-CRAB \tcite{zhang2026crab}               & 184     & \cmark & \cmark & \cmark & \xmark & \xmark & Tests          & \xmark \\
\addlinespace[3pt]
\rowgroup{Security review of pull requests: verdict or report}
SEVRA-Bench \tcite{melo2026sevra}          & ${\sim}$1{,}000 & \pmark & \cmark & \cmark & \xmark & \xmark & LLM        & \pmark \\
MalPR-Bench \tcite{chen2026verdict}        & 89\,+\,50       & \pmark & \cmark & \cmark & \cmark & \pmark & LLM, human & \cmark \\
\midrule
\rowcolor{oursrow}
\textbf{\bench (ours)}                     & \textbf{\numBuggy{}\,+\,\numFixed{}} & \cmark & \cmark & \cmark & \cmark & \pmark & LLM & \cmark \\
\bottomrule
\end{tabular}
\par\smallskip
\begin{minipage}{\linewidth}\scriptsize
Symbols: \cmark~yes, \pmark~partial, \xmark~no.
\emph{Tasks}: vulnerable\,+\,control where both are reported; JitVul counts functions from 879 CVEs.
\emph{Real change}: from project history; \pmark{} for fix-derived changes (reverted or partial patches) or functions from introducing commits.
\emph{No hint}: \pmark{} if hints vary by level (CyberGym) or target metadata is given (SEC-bench Pro).
\emph{Paired control}: fixed or benign controls, not always one per vulnerable input.
\emph{PoC-validated}: \pmark{} if only some cases are (MalPR-Bench: twelve discovery cases; \bench: \S~\ref{sec:verification}).
\end{minipage}
\caption{\textbf{How \bench differs from related benchmarks.} Only \bench uses complete vulnerability-introducing changes (no hint, full repo) and pairs each with its repaired version.}
\label{tab:related}
\end{table*}

The paper makes three main contributions:
\begin{itemize}[leftmargin=*,topsep=2pt,itemsep=1pt,parsep=0pt]
    \item \textbf{Benchmark and the construction pipeline.} \bench, the first comprehensive benchmark for evaluating agentic security review on validated real-world code changes, with \numBuggy{} cases built by an automated, repeatable pipeline that turns fixes into validated pairs of introducing and repaired changes (\S~\ref{sec:construction}).
    \item \textbf{Grading.} A criteria-first judge that credits only root-cause explanations of a target present in the reviewed code, separating correct diagnosis from findings that merely land in the right place, with each verdict's rationale recorded (\S~\ref{sec:evaluation}).
    \item \textbf{Analysis.} Evaluation of eight frontier models and two security-focused harnesses showing where agents diverge, why they miss, and whether harnesses and training cutoffs change recall, with reports, verdicts, and analysis code released (\S~\ref{sec:results}).
\end{itemize}

\section{Related Work}
\label{sec:related-work}

\paragraph{Reviewing code changes.}
Code-review benchmarks range from diff hunks~\citep{li2022codereviewer} to whole PRs with repository context~\citep{zeng2026swrbench,zhang2026aacrbench,kumar2026sweprbench,zhang2026crab}, but their ground truth is review comments across all concerns.
\citet{yu2024security} compare LLMs with static analyzers on security defects from OpenStack and Qt code reviews, and Google's Big Sleep, seeded with a recent SQLite commit, found a loosely related vulnerability before release~\citep{bigsleep2024}.
SEVRA-Bench wraps reverted historical fixes in persuasive PR descriptions~\citep{melo2026sevra}, and \citet{mitropoulos2026contextual} show that such framing biases LLM security review.
MalPR-Bench, the closest design, pairs malicious PRs with benign controls and scores whether a reviewer identifies a rubric-specified target, but derives most PRs from later fixes~\citep{chen2026verdict}.
\bench instead presents the original change that accidentally introduced each vulnerability and was merged before the flaw was reported (Table~\ref{tab:related}).

\paragraph{Vulnerability detection.}
PrimeVul pairs vulnerable and patched functions and shows that detection drops sharply under realistic evaluation~\citep{ding2025primevul}.
\citet{ullah2024llms} find that models sometimes reach correct labels through incorrect reasoning, and \citet{risse2025topscore} argue that function-level labels are ill-posed because vulnerability usually depends on calling context.
JitVul pairs functions from introducing and fixing commits~\citep{yildiz2025jitvul}, and VulnGym evaluates repository-wide audits~\citep{ji2026vulngym}; \bench keeps paired controls and repository access but scores a root-cause explanation of a whole change.

\paragraph{Security agent benchmarks.}
Cybench and BountyBench cover capture-the-flag and bug-bounty tasks~\citep{zhang2025cybench,zhang2025bountybench}, and the reproduction and exploitation benchmarks in Table~\ref{tab:related} require a PoC for a known vulnerability~\citep{lee2025secbench,wang2026cybergym,lee2026secbenchpro}, hint-free discovery and patching~\citep{shi2026cybergyme2e}, or an exploit~\citep{wang2026exploitgym,lee2026exploitbench}.
All score an executable artifact on a whole system, not an explanation of a change.

% Defined here so that Figure 2 lands at the top of the page where Section 3 starts.
% Figure 2: how a SecPRBench case is constructed and verified (four stages with
% artifacts and actors), and what crosses the agent/evaluator boundary.
% Companion to Figure 1 (figs/overview.tex), which shows one case end to end.
% Palette semantics as in Figure 1: blue = agent-visible, amber = evaluator-only;
% the construction agent is drawn with an amber robot, the evaluated agent with
% Figure 1's blue robot. Per-card footers name who performs each stage.
% Preamble requirements (main.tex): fontawesome5; tikz arrows.meta; colours oursrow,
% agentstroke, evalfill, evalstroke, okgreen, badred.
% Geometry (pt, y downward): amber region y 0-118 with four stage cards y 4-114
% (x 12.5-89.5, 96.5-165.5, 172.5-271.5, 280.5-394.5); output band y 121-140.
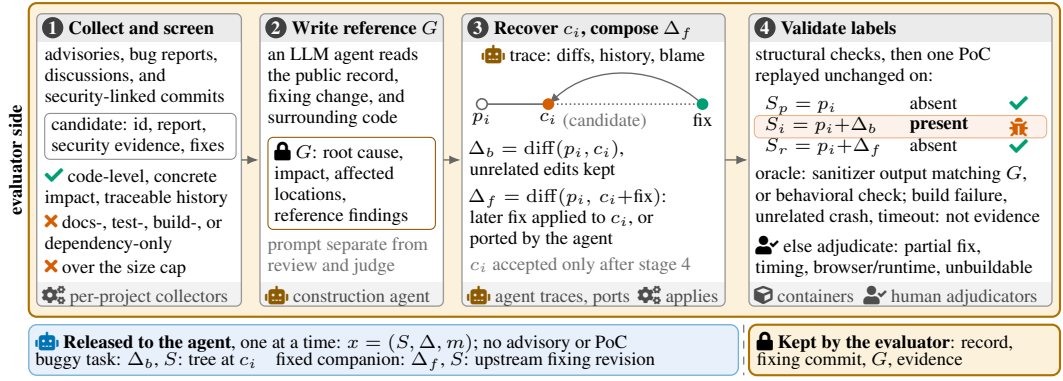
\begin{figure*}[t]
\centering
\begin{tikzpicture}[x=1pt,y=-1pt,font=\scriptsize,
  every node/.style={inner sep=0pt,outer sep=0pt},
  tw/.style={anchor=north west,align=left,execute at begin node={\spaceskip=0pt\xspaceskip=0pt\relax}},
  num/.style={circle,fill=black!75,text=white,minimum size=8pt,font=\scriptsize\bfseries,inner sep=0pt},
  chdr/.style={anchor=north west,inner sep=0pt,minimum height=10.5pt,align=left,text depth=0pt,font=\scriptsize\bfseries},
  foot/.style={anchor=north west,inner xsep=2pt,inner ysep=0pt,minimum height=9pt,align=left,text depth=0pt,text=black!70},
  mchip/.style={anchor=north west,draw=black!45,fill=white,rounded corners=2pt,line width=0.4pt,
                inner xsep=2.5pt,inner ysep=1.5pt,align=left,execute at begin node={\spaceskip=0pt\xspaceskip=0pt\relax}},
  arr/.style={-{Latex[length=1.5mm,width=1.2mm]},line width=0.5pt,black!60},
  ]
% ================================================================ evaluator-side construction region
\node[rotate=90,anchor=center,font=\scriptsize\bfseries] at (4.5,59) {evaluator side};
\draw[fill=evalfill,draw=evalstroke,line width=0.8pt,rounded corners=3pt] (9.5,0) rectangle (397.5,118);
% card frames; header row y 4-14.5, actor footer y 105-114
\foreach \xa/\xb in {12.5/89.5, 96.5/165.5, 172.5/271.5, 280.5/394.5} {
  \begin{scope}
    \clip[rounded corners=2pt] (\xa,4) rectangle (\xb,114);
    \fill[white] (\xa,4) rectangle (\xb,114);
    \fill[black!8] (\xa,4) rectangle (\xb,14.5);
    \fill[black!5] (\xa,105) rectangle (\xb,114);
  \end{scope}
  \draw[draw=black!40,line width=0.5pt,rounded corners=2pt] (\xa,4) rectangle (\xb,114);
}
\draw[arr] (89.5,59) -- (96,59);
\draw[arr] (165.5,59) -- (172,59);
\draw[arr] (271.5,59) -- (280,59);
% ---------------------------------------------------------------- card 1: collect and screen
\node[num] at (18.5,9.25) {1};
\node[chdr] at (24.5,4) {Collect and screen};
\node[tw,text width=71pt] at (15.5,16.5) {advisories, bug reports,\\ discussions, and\\ security-linked commits};
\node[mchip,text width=67pt] at (15.5,41.5) {candidate: id, report,\\ security evidence, fixes};
\node[tw,text width=71pt] at (15.5,62) {\textcolor{okgreen}{\faCheck}\ code-level, concrete\\ impact, traceable history};
\node[tw,text width=71pt] at (15.5,79.5) {\textcolor{badred}{\faTimes}\ docs-, test-, build-, or\\ dependency-only};
\node[tw,text width=71pt] at (15.5,96.5) {\textcolor{badred}{\faTimes}\ over the size cap};
\node[foot] at (12.5,105) {\faCogs\ per-project collectors};
% ---------------------------------------------------------------- card 2: write the reference G
\node[num] at (102.5,9.25) {2};
\node[chdr] at (108.5,4) {Write reference $G$};
\node[tw,text width=63pt] at (99.5,16.5) {an LLM agent reads\\ the public record,\\ fixing change, and\\ surrounding code};
\node[mchip,draw=evalstroke!80!black,text width=58pt] at (99.5,50.5) {\faLock\ $G$: root cause,\\ impact, affected\\ locations,\\ reference findings};
\node[tw,text width=63pt,text=black!55] at (99.5,88) {prompt separate from\\ review and judge};
\node[foot] at (96.5,105) {\textcolor{evalstroke!80!black}{\faRobot}\ construction agent};
% ---------------------------------------------------------------- card 3: recover c_i, compose Delta_f
\node[num] at (178.5,9.25) {3};
\node[chdr] at (184.5,4) {Recover $c_i$, compose $\Delta_f$};
% mini timeline: p_i -- c_i ..... fix, with a backward trace arrow
\coordinate (tp) at (180,38);
\coordinate (tc) at (205,38);
\coordinate (tf) at (263,38);
\draw[black!60,line width=0.6pt] (tp) -- (tc);
\draw[black!60,line width=0.6pt,densely dotted] (tc) -- (tf);
\draw[arr] (tf) to[bend right=42] (tc);
\node[anchor=north,inner sep=0pt] at (222,15.5) {\textcolor{evalstroke!80!black}{\faRobot}\ trace: diffs, history, blame};
\draw[fill=white,draw=black!60,line width=0.6pt] (tp) circle (2pt);
\fill[badred] (tc) circle (2.2pt);
\fill[okgreen] (tf) circle (2.2pt);
\node[anchor=north,inner sep=0pt] at (180,41.5) {$p_i$};
\node[anchor=north,inner sep=0pt] at (205,41.5) {$c_i$};
\node[anchor=north west,inner sep=0pt,text=black!55] at (210.5,41.5) {(candidate)};
\node[anchor=north,inner sep=0pt] at (263,41.5) {fix};
\node[tw,text width=93pt] at (175.5,51.5) {$\Delta_b=\operatorname{diff}(p_i,c_i)$,\\ unrelated edits kept};
\node[tw,text width=93pt] at (175.5,69.5) {$\Delta_f=\operatorname{diff}(p_i,\,c_i{+}\text{fix})$:\\ later fix applied to $c_i$, or\\ ported by the agent};
\node[tw,text=black!55] at (175.5,95.5) {$c_i$ accepted only after stage 4};
\node[foot] at (172.5,105) {\textcolor{evalstroke!80!black}{\faRobot}\ agent traces, ports\ \,\faCogs\ applies};
% ---------------------------------------------------------------- card 4: verify labels
\node[num] at (286.5,9.25) {4};
\node[chdr] at (292.5,4) {Validate labels};
\node[tw,text width=111pt] at (283,16.5) {structural checks, then one PoC\\ replayed unchanged on:};
% S_p / S_i / S_r rows; the S_i row is tinted like Figure 1's "present" pill
\draw[fill=badred!8,draw=badred!50,line width=0.35pt,rounded corners=2pt] (282.5,42.3) rectangle (392.5,50.7);
\node[tw] at (286.5,35) {$S_p=p_i$};
\node[tw] at (286.5,43) {$S_i=p_i{+}\Delta_b$};
\node[tw] at (286.5,51) {$S_r=p_i{+}\Delta_f$};
\node[tw] at (341,35) {absent};
\node[tw] at (341,43) {\textbf{present}};
\node[tw] at (341,51) {absent};
\node[anchor=north,text=okgreen] at (382,35) {\faCheck};
\node[anchor=north,text=badred] at (382,43) {\faBug};
\node[anchor=north,text=okgreen] at (382,51) {\faCheck};
\node[tw,text width=111pt] at (283,61.5) {oracle: sanitizer output matching $G$,\\ or behavioral check; build failure,\\ unrelated crash, timeout: not evidence};
\node[tw,text width=111pt] at (283,87.5) {\faUserCheck\ else adjudicate: partial fix,\\ timing, browser/runtime, unbuildable};
\node[foot] at (280.5,105) {\faCube\ containers\ \ \faUserCheck\ human adjudicators};
% ================================================================ output band: what crosses the boundary
\draw[fill=oursrow,draw=agentstroke!55,line width=0.5pt,rounded corners=3pt] (9.5,121) rectangle (276.5,140);
\draw[fill=evalfill,draw=evalstroke,line width=0.8pt,rounded corners=3pt] (280.5,121) rectangle (397.5,140);
\draw[black!50,dashed,line width=0.5pt] (278.5,121) -- (278.5,140);
\node[tw] at (12,123.5) {\textcolor{agentstroke}{\faRobot}\ \textbf{Released to the agent}, one at a time: $x=(S,\Delta,m)$; no advisory or PoC};
\node[tw] at (12,131.5) {buggy task: $\Delta_b$, $S$: tree at $c_i$\quad fixed companion: $\Delta_f$, $S$: upstream fixing revision};
\node[tw] at (283.5,123.5) {\faLock\ \textbf{Kept by the evaluator}: record,};
\node[tw] at (283.5,131.5) {fixing commit, $G$, evidence};
\end{tikzpicture}
\caption{\textbf{Benchmark construction and validation.} From evidence withheld from evaluated agents, a construction agent writes $G$ and recovers $c_i$; $\Delta_f$ adds the later fix (agent-ported if needed); labels are accepted by mechanical checks, including replay on $S_p$, $S_i$, $S_r$, or by human adjudication.}
\label{fig:pipeline}
\end{figure*}

\paragraph{LLM judges.}
PaperBench and HealthBench judge free-form outputs against per-instance rubrics~\citep{starace2025paperbench,arora2025healthbench}, but judges of code are swayed by response order, identifier names, and misleading comments~\citep{jiang2025codejudgebench}, and c-CRAB and ExploitBench use execution oracles~\citep{zhang2026crab,lee2026exploitbench}.
Execution cannot score an explanation, so our judge derives criteria from a reference, not a human-written rubric, and checks both the target's mechanism and its presence in the reviewed code.

\section{Benchmark Design and Construction}
\label{sec:construction}

Each case is built in four steps from evidence withheld from the evaluated agents (Figure~\ref{fig:pipeline}).

\subsection{Task Formulation}
\label{sec:task}

\bench evaluates whether an agent can identify and explain a vulnerability that a proposed code change introduces.
Each \emph{case} centers on one \emph{target} vulnerability and yields two review tasks: a \emph{buggy task} built from the change that introduced the target (in \emph{partial-fix} cases, a security-relevant change that failed to remove it; \S~\ref{sec:verification}), and a \emph{fixed companion} in which the target is repaired.

\paragraph{Review input.}
The agent receives $x=(S,\Delta,m)$: a repository snapshot $S$, a diff $\Delta$, and change text $m$ from the commit message.
For a buggy task with introducing commit $c_i$ and parent $p_i$, $S$ is the tree at $c_i$ and $\Delta=\Delta_b=\mathrm{diff}(p_i,c_i)$.
For a fixed companion, $\Delta=\Delta_f$ is the introducing change composed with the later repair, again relative to $p_i$; $S$ is the upstream fixing revision, which also contains intervening upstream changes (see Limitations).
The agent may read any code in $S$ but sees no review discussion.
Each task runs in an isolated container with no Git history and with network access limited to the model provider.
The security record, the fixing commit (except code already in a companion's diff or snapshot), the reference, the validation evidence, and the task label stay with the evaluator (Appendix~\ref{app:schema}).

\paragraph{Review output.}
The agent returns a possibly empty set of findings $P(x)$; each should locate the relevant code and explain the trigger, the unsafe operation, and the violated security condition.
A finding detects the target if it explains the causal failure described in the reference $G$ (\S~\ref{sec:reference-construction}) and that failure is present in the reviewed code (\S~\ref{sec:evaluation}).
A fixed companion controls only for its target: $G$ need not cover every vulnerability in the change.
This evaluation set has no training split.

\subsection{Case Construction}
\label{sec:reference-construction}
\label{sec:review-input-construction}

\paragraph{Collection.}
Per-project collectors gather CVE-linked records and public project security reports, including issues without CVE identifiers, with their discussions and fixing commits (Appendix~\ref{app:collectors}).
A candidate is eligible when the evidence establishes a concrete code-level security failure, a plausible trust boundary or untrusted-input path, and a causally connected fixing change.
We keep fixes committed from January 2025 through June 2026, and both review diffs of a case must stay within 30 files and 2,000 added plus deleted lines.

\paragraph{Reference.}
An LLM construction agent, run mainly with Claude Sonnet~5, drafts the reference $G$ from the public security record, the fixing change, and the surrounding code.
$G$ explains the root cause and impact and identifies the code locations needed to understand the causal failure.
We do not audit every reference explanation by hand.
% A construction agent, an LLM agent with access to information withheld from the evaluated agents, writes the reference $G$: the root cause, impact, affected code locations, and reference findings that describe the causal failure. It works from the public security record, which identifies the target and is the primary evidence of its impact and expected failure, and from the fixing change and surrounding code.
%Appendix~\ref{app:prompts} summarizes its instructions, which are separate from the review and judge prompts.

\paragraph{Introducing change.}
The same agent traces the affected code backward from the fix, using diffs, history search, and blame, to a candidate introducing commit $c_i$ with parent $p_i$.
Automatic attribution is unreliable: SZZ-style tools such as V\nobreakdash-SZZ~\citep{bao2022vszz} blame the lines a fix touches but struggle to identify vulnerability-introducing commits~\citep{lu2026vicbench}, and agent-based tracing narrows but does not close this gap on bug-introducing commits~\citep{risse2026bugintroducing}.
We therefore accept an attribution only after validation (as detailed in \S~\ref{sec:verification}), and the buggy task keeps the complete change $\Delta_b$, including unrelated edits.

\paragraph{Fixed companion.}
We apply the later upstream fix to $c_i$ and take the composed change $\Delta_f$ relative to $p_i$; when the fix does not apply cleanly, the construction agent ports it, instructed to keep both the introducing change's functionality and the fix's security-relevant behavior (Appendix~\ref{app:schema}).
Validation decides whether the result removes the target; patch application alone does not.

\subsection{Task Label Validation}
\label{sec:verification}

% Dataset figure (both columns); built by python3 scripts/build_dataset_figure.py.
\begin{figure*}[t]
\centering
\includegraphics[width=\textwidth]{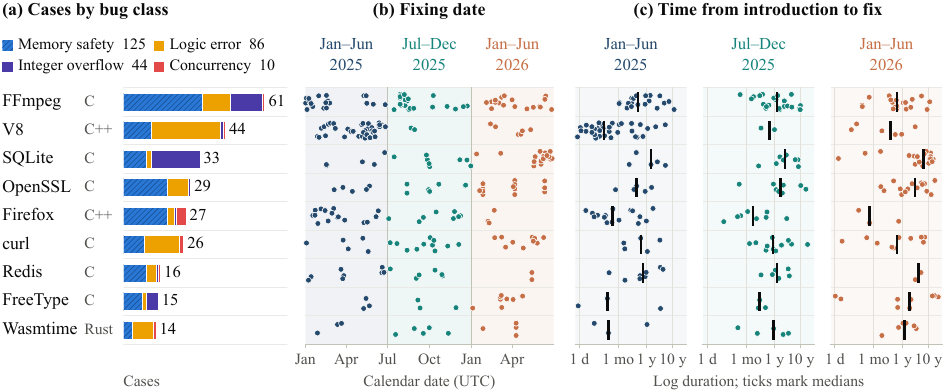}
\caption{\bench by project: (a) bug classes, (b) fixing dates, and (c) time from introduction to fix, grouped by fixing half-year.
Each dot is a case; colors denote fixing half-years in (b) and (c).
Duration panels share a log scale.}
\label{fig:dataset}
\end{figure*}

Validation checks that the introducing change makes the target reachable and that the repair in $\Delta_f$ removes it.
After structural checks, \finalMechanicalVerifiedCases{} cases pass mechanical checks and \finalHumanVerifiedCases{} pass human adjudication; evidence scope varies by case (Appendix~\ref{app:verification}).

\paragraph{PoC replay.}
We build three snapshots in project-specific containers: the parent $S_p$, the buggy tree $S_i=\operatorname{apply}(p_i,\Delta_b)$, and the repaired tree $S_r=\operatorname{apply}(p_i,\Delta_f)$, which differs from the companion snapshot that agents review.
We reuse a public reproducer, adapted to the old revision or harness if needed, or generate one with AnyPoC~\citep{zhao2026anypoc}.
We replay one PoC input and invocation unchanged on all three snapshots and require the target failure to be absent on $S_p$, present on $S_i$, and absent on $S_r$.
The oracle uses sanitizer output matching the reference root cause for memory-safety failures and a behavioral check for supported logic bugs; build failures, unrelated crashes, and timeouts do not count as evidence.

\paragraph{Human adjudication.}
Adjudicators decide the cases that replay cannot settle.
Seven partial-fix cases are a narrow exception to introducing-change tasks: the reviewed repair or hardening leaves the target reachable on both $S_p$ and $S_i$, while replay records its removal on $S_r$.
We retain them only when source evidence directly links the remaining target to the attempted repair in the changed code (Appendix~\ref{app:partial-fixes}).
Other cases need judgment beyond the default oracle, such as interpreting assertions, debugger traces, or races, or cannot be fully replayed on historical builds or platforms.
Adjudication combines the security record, source comparison, fixing change, and available execution evidence; we retain the automated outcome alongside each decision.
% Author follow-up: document who adjudicated, how many adjudicators saw each case, and how disagreements were resolved.

\subsection{Dataset Characteristics}
\label{sec:dataset}

Figure~\ref{fig:dataset} summarizes the \finalVerifiedCases{} cases by project, fixing date, and time from introduction to fix, and Appendix~\ref{app:splits} adds change sizes.
Flaws reported and fixed separately remain separate cases even when they trace to the same change (Appendix~\ref{app:collectors}).

\section{Protocol Grading}
\label{sec:evaluation}

Agents answer a review task with free-text findings, which no execution oracle can check.
Having validated the task labels (\S~\ref{sec:verification}), we now describe how we decide whether an agent found the target.

\subsection{What Counts as a Detection}

We credit a finding when it explains the target's causal failure: what triggers it, which operation becomes unsafe, and which security condition is violated (\S~\ref{sec:task}).
Wording and cited locations may differ from the reference; a sound account of the mechanism suffices, even without an exploit.

We do not credit findings that merely resemble the target.
Some point at the right code but claim a different or generic failure; others describe a real bug next to the target.
The hardest case is a finding that describes the target correctly but in code where it has already been fixed: agents often write nearly identical findings for a change and its fixed companion, naming the same function, operation, and vulnerability class.
Text similarity, location overlap, and Common Weakness Enumeration (CWE) labels cannot separate these cases, so the judge must decide which failure a finding claims and whether it exists in the reviewed code.

In the Wasmtime task of Figure~\ref{fig:overview} (Appendix~\ref{app:schema}), the runtime now expects a defined-table index, but one caller still passes a module-level index.
A finding that ties this mismatch to an access on the wrong table counts as a detection even without a working sandbox escape; a generic bounds warning on the same call does not.

\subsection{Criteria-First Judge}

The judge is \judgeModel{}, with tool use and web access disabled.
It sees the reference root cause, reference findings, and relevant locations, together with the agent's findings and an excerpt of the reviewed change.
For a fixed companion, it receives the original target's reference.
It works through three steps in one session:
\begin{enumerate}[leftmargin=*,topsep=2pt,itemsep=1pt,parsep=0pt]
    \item \textbf{Derive criteria.} It first writes down three to six conditions that separate the target from nearby bugs: the trigger, the unsafe operation, and the missing or incorrect guard, lifetime rule, bound, state transition, or trust boundary.
    \item \textbf{Match the mechanism.} It then checks whether at least one finding meets these criteria, rejecting findings that only point at the same code or describe a neighboring bug.
    \item \textbf{Check the reviewed code.} Finally, it confirms that the causal chain is still present in the reviewed change and that no repair, validation, bounds check, or synchronization breaks it; matching the historical reference alone does not show that the vulnerability is live.
\end{enumerate}
The judge returns a single verdict per task, along with its criteria, confidence, and reasoning.
Appendix~\ref{app:prompts} summarizes the prompt, and Appendix~\ref{app:scoring} gives implementation details, including how change excerpts are selected.

\subsection{Judge Reliability}

Across \numJudgeStabilityReports{} nonempty buggy reports, Opus~5.5 and Sonnet~4.6 agree on \judgeStabilityAgreement{} of verdicts.
In a separate manual check, both judges matched our adjudications on all 25 randomly sampled agreement controls.
Among 25 selected disagreements, our adjudications supported Opus in 21 cases and Sonnet in four.
These checks demonstrate strong agreement between judges and support Opus on most audited disagreements.
The diagnostic sample does not estimate overall judge accuracy.
Appendix~\ref{app:judge-audit} gives the audit and grading examples.

\subsection{Metrics}

\paragraph{Target recall.}
Our main metric is the fraction of buggy tasks in which the judge credits a detection.
With $d_i\in\{0,1\}$ as the verdict for buggy task $i$, target recall is
\begin{equation}
 R = \frac{1}{N}\sum_{i=1}^{N} d_i,
\end{equation}
where $N=\numBuggy{}$ is the number of buggy tasks, the same for every agent.
A task counts as missed if the agent returns nothing or its run ends in an error.

\paragraph{Hit@1.}
Hit@1 is the fraction of buggy tasks whose first finding detects the target; it models a developer who reads only the top finding.
When the judge's rationale does not identify the credited finding, we report bounds.

\paragraph{Report volume and cost.}
\looseness=-1
We also count findings per task, since each one is something a developer must read.
Findings that match no target are not necessarily false positives: the reference does not list every vulnerability in a change, and several findings may describe the same defect.
Cost excludes construction and judging (Appendix~\ref{app:results-accounting}).

% Current numbers derive from data/results_265_eight_models_opus55_high.json via scripts/build_265_results.py.
\section{Evaluation}
\label{sec:results}

On the \numBuggy{} buggy changes and their fixed companions, we ask how often agents find the introduced flaw (RQ1), why they miss (RQ2), whether a security-focused harness helps (RQ3), and how they perform on fixed versions (RQ4).

\begin{table*}[t]
\centering\footnotesize
\setlength{\tabcolsep}{3.3pt}

\begin{tabular}{@{}lrrrrrrrrrr@{}}
\toprule
& \multicolumn{3}{c}{\textbf{Detection}} & \multicolumn{2}{c}{\textbf{Misses}} & \multicolumn{2}{c}{\textbf{Fixed companion}} & \multicolumn{3}{c}{\textbf{Effort}} \\
\cmidrule(lr){2-4}\cmidrule(lr){5-6}\cmidrule(lr){7-8}\cmidrule(l){9-11}
Agent & Found & Recall & Hit@1 & Rej. & Empty & Flagged & $\Delta$ Findings & Tok (k) & Est. cost (\$) & Min. \\
\midrule
% Generated from data/results_265_eight_models_opus55_high.json by scripts/build_265_results.py.
\rowgroupn{11}{\claudecodeicon\ Claude Code}
Grok~4.6 & 207 & \heat{40}{\textbf{78.1}} & \heat{37}{46.4--55.5} & 53 & 5 & 0 & \dropcell{4} & 43.3 & 3.20 & 12.6 \\
GLM~5.3 & 198 & \heat{38}{74.7} & \heat{40}{\textbf{49.8--58.1}} & 66 & 1 & 0 & \dropcell{6} & 72.6 & 4.99 & 20.8 \\
Claude Opus~5.5 & 196 & \heat{37}{74.0} & \heat{37}{47.2--55.1} & 63 & 6 & 0 & \dropcell{13} & 7.2 & 0.44 & 1.4 \\
Claude Sonnet~5 & 157 & \heat{28}{59.2} & \heat{31}{41.1--47.9} & 100 & 8 & 0 & \dropcell{6} & 31.9 & 0.83 & 5.5 \\
MiniMax~M3 & 86 & \heat{10}{32.5} & \heat{10}{20.4--26.8} & 126 & 53 & 0 & \risecell{11} & 38.5 & 3.33 & 11.5 \\
\addlinespace[3pt]
\rowgroupn{11}{\codexicon\ Codex CLI}
GPT-6 sol & 202 & \heat{39}{76.2} & \heat{37}{44.9--57.0} & 49 & 14 & 0 & \dropcell{14} & 15.7 & 0.60 & 3.3 \\
GPT-5.6 sol & 148 & \heat{25}{55.8} & \heat{31}{43.8--46.4} & 76 & 41 & 0 & \dropcell{31} & 9.3 & 0.74 & 2.7 \\
GPT-6 luna & 147 & \heat{25}{55.5} & \heat{31}{42.6--46.4} & 76 & 42 & 0 & \dropcell{28} & 14.8 & 0.02 & 2.7 \\
GPT-5.6 sol + Codex Security & 126 & \heat{20}{47.5} & \heat{26}{38.5--41.9} & 67 & 72 & 0 & \dropcell{25} & 31.3 & 3.74 & 7.3 \\
\midrule
Any agent (oracle) & 242 & 91.3 & --- & --- & --- & --- & --- & --- & --- & --- \\
\bottomrule
\end{tabular}
\caption{Results on \numBuggy{} tasks (rates in \%); darker is higher within a column.
Hit@1: first finding detects the target; unresolved ranks count as none or all first.
Rej./Empty: rejected/empty reports; Flagged: positive target verdicts.
$\Delta$ Findings: change in mean findings per scan from the buggy change to its fixed companion ($\downarrow$ fewer, $\uparrow$ more).
Effort: mean output tokens, estimated cost, and minutes per scan.}
\label{tab:main}\label{tab:cost}
\end{table*}

\paragraph{Setup.}
We evaluate Claude Opus~5.5, Claude Sonnet~5, Grok~4.6~\citep{xai2026grok46}, GPT-5.6 sol~\citep{openai2026gpt56}, GPT-6 sol, GPT-6 luna, GLM~5.3~\citep{zai2026glm53,glm5team2026glm5}, and MiniMax~M3~\citep{minimax2026m3} on the same inputs, withholding the reference and the later fix (Appendix~\ref{app:prompts}).
Each model runs in an agent harness, the program that provides its tools and drives its loop.
For direct review, the three GPT agents run in the Codex CLI~\citep{openai2026codexcli} at \code{xhigh} reasoning effort, and the others in Claude Code~\citep{anthropic2026claudecode}, both general-purpose coding harnesses.
For RQ3, we compare two security-focused harnesses: GPT-5.6 sol also runs as a static diff scan inside OpenAI's Codex Security plugin~\citep{openai2026codexSecurityDocs}, and Opus~5.5 runs inside Anthropic's Defending Code reference harness~\citep{anthropicDefendingCodeHarness}.
Web tools are disabled, and runs have no turn or token limit but time out after an hour (Appendix~\ref{app:agent-config}).

\subsection{RQ1: How Often Do Agents Find the Introduced Vulnerability?}

% Section 5 overview figure. Generated PDF: figs/results_overview.pdf, built by
%   python3 scripts/build_265_results.py
% from data/results_265_eight_models_opus55_high.json; --check cross-checks the derived counts.
% Drawn at print size (6.3 in = \textwidth), so its 7 pt text prints at 7 pt.
% Place with \input{figs/results_overview} near the start of Section 5 (e.g. right after the first RQ heading).
\begin{figure*}[t]
\centering
\includegraphics[width=\textwidth]{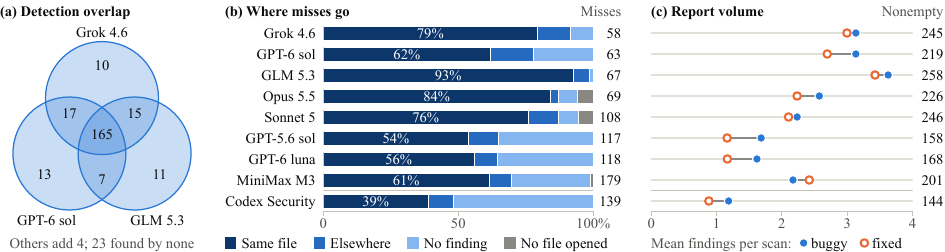}
\caption{Results on the \numBuggy{} buggy tasks and fixed companions.
(a)~Targets found by the three leading agents.
(b)~Misses: a finding in a root-cause file for another mechanism (\%), findings only elsewhere, no finding, or no root-cause file opened.
(c)~Mean findings per buggy and fixed scan; Nonempty: fixed companions with any finding.}
\label{fig:results}
\end{figure*}

Strong agents tie on easy targets and separate on contested ones.
Grok~4.6, GPT-6 sol, GLM~5.3, and Opus~5.5 lead, each finding about three in four targets (Table~\ref{tab:main}).
Grok is highest at \recallHigh{}, but no pair among the four differs significantly (Appendix~\ref{app:pairs}); they differ more in speed, and Opus~5.5 is the fastest of all eight.
To see where agents separate, we group targets by how many of the other seven agents find them.
On easy targets, which nearly all other agents find, every agent but MiniMax~M3 finds at least nine in ten; on contested ones, which only a few others find, the top four still find about half, while the other four find at most one in five (Appendix~\ref{app:pairs}).
A newer model can close that gap: with the same harness and reasoning effort, GPT-6 sol finds 58 targets that GPT-5.6 sol misses and misses only four that it finds, with the largest gains on integer overflows and logic errors (Table~\ref{tab:class}).
As in prior security benchmarks~\citep{wang2026cybergym,lee2026secbenchpro,wang2026exploitgym}, combining agents improves coverage, but here the gains shrink quickly: the three leading agents together find about nine in ten targets, and all eight find only one more than the top four (Figure~\ref{fig:results}a).
Even combined, the eight agents leave \sNeverFound{} targets unfound, far more than independent errors would predict (Appendix~\ref{app:never-found}).
Training cutoffs make no clear difference: adjusted post-cutoff recall is several points lower for four of five testable models and slightly higher for GPT-6 sol, but every interval spans zero, and fixing dates cannot show training exposure (Appendix~\ref{app:temporal}).

\subsection{RQ2: Why Do Agents Miss?}
Most misses are misdiagnoses, not oversights.
In 68\% of scanned direct-review misses, the agent files a finding in a root-cause file, but the judge rejects it as describing a neighboring mechanism (Figure~\ref{fig:results}b); these reports mostly argue a competing diagnosis rather than declaring the site safe, and we did not re-adjudicate them.
In curl CVE-2026-3805, a nine-line change to \code{lib/smb.c} leaves a request's path pointing into a buffer that dangles once its connection is freed; all eight agents report on that file, but four call the buffer leaked and the GPT agents call the line a logic bug, and our audit upholds one such rejection (Appendix~\ref{app:judge-audit}).
Agents rarely fail to reach the right file, but only because the diff almost always includes it.
Nor is effort lacking: within difficulty strata, Opus's token spend does not differ significantly between misses and hits (Appendix~\ref{app:miss-anatomy}).
Logic errors are harder than memory-safety flaws: averaged over agents, recall is about half against nearly three in four, and logic errors make up most targets that every agent misses (Table~\ref{tab:class}); bug class is confounded with project, but the gap persists outside V8.
How far the mechanism spreads also matters: pooled recall falls from about seven in ten with one root-cause file to under four in ten with five or more, even among diffs of similar size, whereas change size and age show weak or no association; these splits are observational (Appendices~\ref{app:miss-anatomy} and~\ref{app:splits}).

\subsection{RQ3: Does a Security-Focused Harness Help?}
With GPT-5.6 sol at the same reasoning effort, Codex Security lowers recall by 8.3 points (Table~\ref{tab:main}), mostly by abstaining: on most tasks found only by direct review, it returns no finding (Appendix~\ref{app:results-codex}).
It also files fewer findings (Figure~\ref{fig:results}c), despite about five times the mean estimated cost.
Anthropic's Defending Code reference harness~\citep{anthropicDefendingCodeHarness} raises Opus~5.5's recall by 7.7 points on the pairs it completed, all but five, with more findings at about ten times the median nominal cost (Appendix~\ref{app:results-anthropic-harness}).
It alone finds all ten concurrency targets, where no direct-review agent finds more than six, and it finds six of the 22 unfound targets it completed, through breadth rather than ranking: each credited report there carries over a dozen findings.
Both comparisons fix the backend model and reasoning effort; they remain descriptive because prompts, budgets, and operating points differ.

\subsection{RQ4: How Do Agents Perform on Fixed Versions?}
\label{sec:fixed-results}
No fixed companion receives a positive target verdict, in direct review or either harness (Table~\ref{tab:main}; Appendix~\ref{app:results-anthropic-harness}).
This is a narrow check: a negative verdict does not show whether a report reasserts the removed flaw, so these verdicts measure neither false-positive rates nor report precision.
Location overlap alone would credit about three in four companion scans: over nine in ten nonempty reports still name a root-cause file.
Still, every direct-review agent but MiniMax~M3 files fewer findings after the repair ($\Delta$ Findings in Table~\ref{tab:main}; Figure~\ref{fig:results}c), and the drop follows correct diagnoses: when an agent's only finding on a buggy change is credited, its companion report is empty almost half the time, against about one in six when that finding is rejected.
A fixed companion is thus a control only for its target, not a claim that the whole change is safe (Appendix~\ref{app:companions}).

\section{Conclusion}

\looseness=-1
\bench evaluates vulnerability diagnosis on real code changes, pairing \numBuggy{} buggy tasks with repaired companions.
Recall across eight direct-review agents ranges from \recallLow{} to \recallHigh{}, with four agents statistically tied at the top.
Strong agents tie on easy targets but separate on contested ones.
Even combined, the eight agents leave \sNeverFound{} targets unfound.
Most misses are misdiagnoses despite findings in the flawed file.
Security-focused harnesses move recall in both directions: Codex Security abstains more, while Anthropic's Defending Code harness finds more at higher cost.
On repaired changes, the judge never rules the removed flaw still present, though this does not establish report precision.
The automated pipeline supports refreshes from new disclosures; released tasks, reports, and verdicts let others inspect and re-grade the findings.
Better diagnosis is the key remaining challenge.

\FloatBarrier
\label{main-text-end}
\section*{Limitations}
\label{sec:limitations}

\paragraph{Scope.}
\bench covers publicly disclosed vulnerabilities in \numProjects{} widely deployed C, C++, and Rust projects, with changes of up to 30 files and 2,000 changed lines (\S~\ref{sec:reference-construction}), so our conclusions concern security review of systems code.
Its targets were later disclosed and fixed, so they may differ from flaws that remain undiscovered.
Because every buggy task contains a validated target, \bench measures whether agents find and diagnose known vulnerabilities, not how often ordinary pull requests are vulnerable.
Fixed companions are reviewed at the upstream fixing revision, which can include intervening changes, and control only for their targets, so \bench does not measure report precision (\S~\ref{sec:fixed-results}).

\paragraph{Labels and judging.}
Of the \numBuggy{} cases, \finalMechanicalVerifiedCases{} pass mechanical checks, mainly PoC replay on the parent, introducing, and repaired trees (\S~\ref{sec:verification}).
The other \finalHumanVerifiedCases{}, such as partial fixes and races, rest on human adjudication of source and execution evidence, and we keep the automated outcome beside each decision.
Our judge, \judgeModel{}, is also an evaluated model, as is Sonnet~5, which drafted most references (\S~\ref{sec:reference-construction}).
Switching the judge from Sonnet~4.6 to Opus~5.5 changes Opus~5.5's own verdicts on only seven tasks, four up and three down, and the two judges agree on \judgeStabilityAgreement{} of \numJudgeStabilityReports{} nonempty buggy reports (Appendix~\ref{app:judge-fields}).
Our manual audit supports the judge on all 25 sampled agreements and 21 of 25 disagreements (Appendix~\ref{app:judge-audit}), and we release reports and verdicts so that other judges can re-grade them.

\paragraph{Runs and comparisons.}
Each agent has one run per task, without matched token budgets, so we report tokens, cost, and time beside recall (Table~\ref{tab:main}); each result describes a model in its harness, not the model alone.
We do not repeat runs; paired tests over all \numBuggy{} tasks separate each of the four leading agents from each of the other four (Appendix~\ref{app:pairs}).
Harness comparisons fix the backend model and reasoning effort but not prompts or budgets, so they compare each harness as a whole, in the configuration we ran (Appendix~\ref{app:results-codex}).

\paragraph{Training data.}
Public security records and fixes may appear in model training data.
Agents review buggy changes without the security record, the fixing commit, Git history, or web tools (\S~\ref{sec:task}; Appendix~\ref{app:agent-config}), and harness comparisons hold the backend model, and so its training data, fixed.
For the five models with fixes on both sides of their documented cutoffs, adjusted recall after the cutoff is lower for four of them, but no difference is statistically clear (Appendix~\ref{app:temporal}).
Because collection, tracing, and PoC replay are automated, \bench can be refreshed with vulnerabilities disclosed after future training cutoffs.

\section*{Ethics Statement}
The tasks are based on public security records and linked fixes.
We reuse, adapt, or generate PoCs for these historical vulnerabilities only to validate task labels; the benchmark itself evaluates defensive source review.
All nine projects are open source, and their code stays under its upstream licenses, such as MPL-2.0 for Firefox, BSD-3-Clause for V8, and Apache-2.0 for OpenSSL and Wasmtime.
Tasks contain only public code, commit messages, and security records; we collect no personal data beyond developer names that the projects already publish.
Benchmark labels do not establish a new vulnerability in a current release; any new finding requires independent validation and responsible disclosure to the affected project.

An AI coding assistant helped inspect artifacts, compute descriptive statistics, and edit and proofread the manuscript.
Language model agents were also used for task construction, PoC preparation, evidence interpretation, and scoring.
Agent validation reports do not count as independent human judgments.
The authors are responsible for the data, adjudication decisions, and claims.

% acl.sty selects the official acl_natbib bibliography style.
\bibliography{references}
\FloatBarrier
\clearpage
\appendix
% Frozen October 10 release admission counts.
\newcommand{\funnelMechanicalVerifiedTasks}{\finalMechanicalVerifiedCases}
\newcommand{\funnelHumanVerifiedTasks}{\finalHumanVerifiedCases}

\section{Collector Inputs and Inclusion Evidence}
\label{app:collectors}

\paragraph{Construction-stage counts.}
Table~\ref{tab:construction-funnel} summarizes collection, selection, and validation.
Counts before selection pool successive collection rounds; the selected cases form the evaluation population (Appendix~\ref{app:verification}).
Each selected case has a buggy task and a fixed companion.
Targets follow the collected security records: flaws reported and fixed separately are retained as separate cases, each with its own reference and fixed companion, even when they trace to the same introducing change.

% Source: ../tmp/pipeline_funnel_report.md (2026-09-22), stages 0, 1, and 4.
\begin{table*}[t]
\centering\small

\begin{tabular}{@{}lr@{}}
\toprule
Stage & Count \\
\midrule
Raw source records, all rounds & $>$7,000 \\
Triaged candidates & 768 \\
Selected cases & \numBuggy{} \\
\midrule
Validated cases: mechanical & \funnelMechanicalVerifiedTasks{} \\
Validated cases: human adjudication & \funnelHumanVerifiedTasks{} \\
\bottomrule
\end{tabular}
\caption{Collection and validation counts.
Raw records include overlapping reports and repeated records of the same issue.
Selected cases are accepted through executable checks (mechanical) or additional human adjudication.}
\label{tab:construction-funnel}
\end{table*}

\paragraph{Collection and eligibility.}
We collect candidates from two complementary sources: CVE-linked vulnerability records and public project security reports, including issues without assigned CVE identifiers.
Project-specific collectors gather the associated advisories, bug reports, discussions, and fixing commits, and normalize them into candidate records containing an issue identifier, available report text, evidence of security relevance, and candidate fixing revisions.
Beyond the eligibility criteria in \S~\ref{sec:reference-construction}, we exclude documentation-only, test-only, dependency-update, and build-metadata changes, as well as unsupported security claims.
Introducing and fixing revisions must lie on the project's mainline, and fixing revisions have committer timestamps on or after January 1, 2025, and before July 1, 2026.
Table~\ref{tab:collectors} records the source-specific inputs used by the project collectors.
A security label is an inclusion signal, not a common severity scale across projects.
Cases without CVE identifiers use project issue identifiers or publicly documented security-linked fixing commits.

\begin{table*}[t]
\centering\small

\begin{tabular}{@{}p{0.13\linewidth}p{0.39\linewidth}p{0.42\linewidth}@{}}
\toprule
Project & Collector input & Security signal \\
\midrule
curl & Advisory index and machine-readable records & Project-maintained advisory information \\
FFmpeg & Trac, advisories, CVEs, and Git history & Advisory or commit evidence of a concrete security failure \\
Firefox & Bugzilla records and comments & Security-bug flags and critical/high/moderate security labels \\
FreeType & OSS-Fuzz-linked commits and GitLab issues & Source-based evidence of memory errors or malicious-font crashes \\
OpenSSL & OpenSSL advisory data and GitHub issues/PRs & Upstream ratings and source-based security screening \\
Redis & GitHub Security Advisories and OSV & Published advisories with security-fix references \\
SQLite & Project CVE page, Fossil forum, and tracker & Input-reachable memory-safety or corruption evidence \\
V8 & Chromium security issues fixed in V8 & Chromium security labels, restricted to fixes in V8 \\
Wasmtime & GitHub Security Advisories and OSV & Published advisories linked to fixes \\
\bottomrule
\end{tabular}
\caption{Sources and security-relevance signals used by each project's collectors.
All projects additionally require a traceable code change suitable for benchmark task construction.}
\label{tab:collectors}
\end{table*}

The shared size thresholds are 30 changed files and 2,000 added plus deleted lines, with no minimum-line threshold.
All \numBuggy{} buggy changes and their fixed companions satisfy these caps.
We retain the introducing and fixing revisions and the source evidence so that eligibility can be checked independently of agent results.

\section{Validation Evidence and Adjudication}
\label{app:verification}

\paragraph{Recorded outcomes and population.}
Of the \numBuggy{} cases, identified by project and bug identifier, \finalMechanicalVerifiedCases{} are accepted through the mechanical route and \finalHumanVerifiedCases{} through human adjudication (Table~\ref{tab:verification}).
For the human-adjudicated cases, the available execution evidence and control coverage vary by case.
Adjudication also distinguishes target failures from safe input rejection, feature absence, and unrelated execution failures in control runs.

\begin{table*}[t]
\centering\small

\begin{tabular}{@{}lrrr@{}}
\toprule
Project & Cases & Mechanical & Human \\
\midrule
% Generated from data/results_265_eight_models_opus55_high.json by scripts/build_265_results.py.
curl & 26 & 23 & 3 \\
FFmpeg & 61 & 53 & 8 \\
Firefox & 27 & 18 & 9 \\
FreeType & 15 & 12 & 3 \\
OpenSSL & 29 & 28 & 1 \\
Redis & 16 & 12 & 4 \\
SQLite & 33 & 29 & 4 \\
V8 & 44 & 32 & 12 \\
Wasmtime & 14 & 10 & 4 \\
\midrule
Total & 265 & 217 & 48 \\
\bottomrule
\end{tabular}
\caption{Validation routes for the \numBuggy{} evaluated cases.
Mechanical: cases accepted through recorded executable checks; Human: additional cases accepted by adjudication.
Each case contributes two review inputs.}
\label{tab:verification}
\end{table*}

Evidence scope also varies within a route.
The FFmpeg case fixed by commit 4eca335d94ca passes mechanical validation with a source-based check of the repair instead of a replay on the repaired tree, and some Firefox cases replay their PoCs in component test harnesses.
Per-case validation records, which we release, document the evidence behind each route decision.

\paragraph{Structural and mechanical checks.}
Before replay, structural checks cover commit hashes, task metadata, diff structure, and patch applicability.
For mechanical validation, the shared validator reconstructs three snapshots from the introducing revision's first parent: the parent $S_p$, the buggy tree $S_i$, and the repaired tree $S_r$.
When no public reproducer exists, AnyPoC generates one from the vulnerability report, the fix, and the repository.
The validator records build results, patch-application status, exit codes, sanitizer output, and execution logs.
For introducing changes, validation requires the same PoC and invocation to show the target failure on $S_i$ and no target failure on $S_p$ or $S_r$.
The observed failure must match the reference root cause; unrelated crashes, build failures, and timeouts do not establish a positive verdict.
We inspect the recorded evidence to confirm the observed behavior, not just build or crash status.

\paragraph{Partial fixes.}
\label{app:partial-fixes}
We distinguish partial fixes in the reviewed change from those in its companion.
A reviewed partial fix is a narrow exception to the requirement that a buggy task introduce the target: the same flaw may remain reachable before and after the change.
Real repair commits can leave vulnerabilities unresolved; reviewers should identify why the changed logic fails to remove the target.
We retain such cases only when source evidence establishes this causal link and the companion removes the target; an unrelated pre-existing flaw does not qualify.
A companion that still contains the target is not a valid repaired control; one that removes it may still contain another defect.

In the SQLite case fixed by commit 03380a55a521, an attempted zipfile repair uses a 30-byte header constant in a bounds check that requires 46 bytes.
The same PoC triggers the target overread in both the parent and reviewed versions.
The companion corrects the constant and safely rejects the malformed archive.

\paragraph{Failures not observable with AddressSanitizer (ASan).}
In Firefox issue 1948111, the introducing change adds repeated locale-string writes that can race with worker reads.
A recorded browser run exercises the relevant paths without an ASan error; source comparison with the fix identifies the missing synchronization.
This motivates race-aware or source-based adjudication, rather than interpreting sanitizer silence as absence of the bug.
FreeType OSS-Fuzz issue 499455833 offers a complementary example: a debugger shows a point index outside the glyph's logical zone, while historical over-allocation masks the access from ASan.
Human adjudication reads this evidence alongside the source and repair.

\paragraph{Dataset and control identity.}
The released snapshot records the selected cases and the hashes, commit identifiers, and diff statistics of all \finalVerifiedInputs{} review inputs.
Per-case validation records identify the source trees and controls that each check exercised.
Each validation verdict applies to the recorded companion patch, not to other patches for the same bug.

\section{Task Schema and Snapshot Contract}
\label{app:schema}

\paragraph{Construction implementation.}
Collectors produce normalized source records from public security evidence.
The construction agent writes reference findings and identifies fixing and introducing revisions.
A mechanical wrapper then extracts the introducing change and packages the buggy task.
The companion generator first attempts mechanical fix composition and uses a construction agent when adaptation is needed.
Relevant-code annotations enrich the evaluation reference.
The benchmark loader removes the reference and other evaluator-only metadata before review; the review prompt presents the commit message and file diffs, with repository context available through tools.

\paragraph{Stored fields.}
Task records retain repository and revision identities, the review change, and the evaluation reference.
Table~\ref{tab:info-boundary} distinguishes agent-visible inputs from constructor and evaluator metadata; the released schema documents individual fields.

\begin{table*}[t]
\centering\small

\begin{tabular}{@{}p{0.27\linewidth}p{0.29\linewidth}p{0.38\linewidth}@{}}
\toprule
Information & Agent & Constructor/evaluator \\
\midrule
Review diff, commit message & Yes & Yes \\
Task source snapshot & Yes & Yes, with exact revision recorded \\
Target security record & No & Yes \\
Later fixing commit/diff & No, except in a fixed companion & Yes \\
Reference findings and annotated root-cause locations & No & Yes \\
Buggy/fixed label and case identifier (e.g., CVE or bug ID) & No & Yes \\
Future Git history & No & Constructor only \\
\bottomrule
\end{tabular}
\caption{Information boundary.
Reference information is withheld from the reviewing agent.}
\label{tab:info-boundary}
\end{table*}

\paragraph{Task materialization.}
\looseness=-1
The setup tool exports each task's recorded source snapshot using \code{git archive} and provides the review diff separately.
Snapshots use the introducing revision for buggy tasks and the recorded fixing revision for companions.
The recorded parent identifies the base of the review diff, rather than the exported snapshot.
The exported workspace does not contain Git history.
Each task runs in its own isolated Docker container; evaluator-only artifacts remain outside it, and outbound network access is restricted to configured provider endpoints.
Removing evaluator-only labels, references, and validation metadata preserves the upstream change text, diff, and source files.

\paragraph{Paired materialization for validation.}
Let $\Delta_b=\operatorname{diff}(p_i,c_i)$ and let $\Delta_f$ be the composed fixed patch.
The shared validator materializes $S_i=\operatorname{apply}(p_i,\Delta_b)$ and $S_r=\operatorname{apply}(p_i,\Delta_f)$; the first is the buggy-task snapshot.
Validation reconstructs both changes from their common parent, whereas agents review the recorded fixing snapshot for companions.
Per-case records identify whether controls use an upstream fixing revision or an adapted source tree (Appendix~\ref{app:verification}).
A structural check confirms patch applicability, which alone does not establish target removal.

\paragraph{A concrete target.}
\looseness=-1
Wasmtime CVE-2026-34946 illustrates the need for repository context: the changed \code{table.fill} runtime expects a defined-table index, but the unchanged Winch caller still supplies a module-level index and caller context.
For imported tables, these can identify different tables.
A useful report must connect this interface mismatch to the unchanged caller, rather than merely warn about bounds.

\section{Prompt Design and Summaries}
\label{app:prompts}

We summarize the substantive instructions and input fields below.
These are condensed descriptions of the prompts, not verbatim copies of the original templates.
Runtime paths, tool schemas, and repeated formatting instructions are omitted.

\paragraph{Direct-review prompt.}
For the Claude Code agents, the system message sets the security-review role, repository-investigation workflow, and reporting policy.
The task message supplies the commit message and each changed file's path, status, and diff.
Source code is available through file-reading, search, and shell tools.
The core instructions are:

\begin{quote}\small
Review the proposed change for security failures.
Begin with changed functions and follow relevant callers, callees, and data structures.
Check whether the change breaks assumptions about types, bounds, ownership, error handling, or control flow, including exceptional paths.
Read the code; do not infer safety from stated intent.

Report each plausible failure with a concrete location, a path to it, and its security impact.
State uncertainty; a complete exploit is not required.
Exclude unsupported speculation, style issues, and failures demonstrably prevented by the code.
Continue reviewing after each finding.
\end{quote}

In Claude Code, each finding is submitted through our reporting tool as JSON with a title, explanation, location, bug type, severity, and confidence; related locations and suggested fixes can supplement it.
This reporting policy favors candidate discovery.
The benchmark's detection criterion is applied afterward by the judge.

\looseness=-1
The three GPT agents' Codex CLI prompt supplies the same commit message and diff, requests security issues introduced or made reachable by the change, limits findings to changed files, and requires a final JSON report.
Codex Security and Anthropic's Defending Code reference harness follow their own review workflows (Appendix~\ref{app:results-codex}).

\paragraph{Criteria-first judge prompt.}
The judge receives the task type, the reference root cause, reference findings, relevant locations, the agent's findings, and a selected change excerpt.
These reference fields are withheld from the reviewing agent.
Fixed companions use the original target reference together with the evaluated companion's change.
The core instructions are:

\begin{quote}\small
Derive three to six necessary criteria for the reference vulnerability before assessing the reports.
Make the criteria specific enough to distinguish neighboring bugs.
Check whether at least one finding identifies the same trigger, unsafe operation, and violated guard, ownership rule, bound, state transition, or trust boundary.

Credit detection only if the change evidence supports that causal chain and no repair or mitigation breaks it.
Reject adjacent bugs, location-only matches, hypothetical future risks, and historical or already-fixed failures.
Examine all findings and return one task-level verdict.
\end{quote}

The JSON response records detection, criteria, a match score, confidence, current-code support, mitigation flags, and a short explanation.
Appendix~\ref{app:scoring} covers excerpt selection and why this verdict does not measure target alarms on fixed companions.

\paragraph{Construction instructions.}
Construction agents use public security evidence and fixing history to identify the root cause, affected code, and introducing and fixing revisions.
When mechanical repair composition fails, a construction agent adapts the repair to the older source while preserving both the intended functionality of the introducing change and the fix's security-relevant behavior.
The validation pipeline then checks these artifacts.

\section{Scoring Details and Their Limits}
\label{app:scoring}

\paragraph{Criteria-first evaluation.}
Our evaluation runner prompts the judge to derive task-specific criteria from the reference root cause and reference findings.
The judge evaluates the agent's final report against these criteria, using relevant diff sections selected by affected files and falling back to the full diff when necessary.
Excerpts longer than 30,000 characters are truncated.
The judge returns a structured verdict indicating detection, confidence, and supporting reasoning.
Tasks with empty reports are counted as non-detections without a judge call, and all tasks remain in the recall denominator.

\paragraph{Manual audit.}
\label{app:judge-audit}
\looseness=-1
We adjudicated \numJudgeAudit{} evaluation records with LLM assistance and checked each adjudication manually.
The sample contains 25 selected Sonnet--Opus disagreements and 25 randomly selected agreement controls with nonempty reports.
We compared the reference, findings, verdicts, and code, including the supplied change excerpt where available.

\begin{table}[t]
\centering\small
\setlength{\tabcolsep}{4pt}
\begin{tabular}{@{}lrr@{}}
\toprule
Audited group & Opus~5.5 & Sonnet~4.6 \\
\midrule
% Generated by scripts/build_judge_audit.py from the frozen author audit.
Judge disagreements & 21/25 & 4/25 \\
Agreement controls & 25/25 & 25/25 \\
\bottomrule
\end{tabular}
\caption{Agreement with our manual adjudications.}
\label{tab:judge-audit}
\end{table}

\looseness=-1
Disagreements were deliberately oversampled, so Table~\ref{tab:judge-audit} reports each group separately; pooled percentages do not estimate overall judge accuracy.
In the four cases where Opus departs from our adjudication, it broadened a target trigger or sink, supplied a missing causal step, or combined partial findings across mechanisms.
Agreement concerns the binary verdict, not every supporting claim: one correct negative included a repair rationale whose decisive code lay outside the judge's excerpt.

\paragraph{Matching the mechanism despite different wording.}
\looseness=-1
In curl CVE-2026-9546, clearing \code{CURLOPT\_REFERER} with automatic Referer handling disabled leaves \code{state.referer} unchanged: the new conditional update has no clearing branch.
The audited report identifies this transition and subsequent header use, but describes a dangling pointer and cites the setter.
Opus credits the shared causal chain, consistent with our adjudication; Sonnet rejects it over classification and location differences.
The accepted match concerns the missing reset; we did not independently validate the report's stronger claim about memory lifetime.

\paragraph{Separating neighboring mechanisms.}
\looseness=-1
In the OpenSSL CVE-2025-9232 companion, the reference concerns an unterminated \code{host} buffer reaching \code{strstr}; the fix adds \code{host[sl] = \textquotesingle\textbackslash0\textquotesingle;}.
The audited report instead alleges that the bracketed host \code{[]} yields an empty, terminated search string whose repeated matches advance past the end of \code{no\_proxy}.
Both judges reject target detection, consistent with our adjudication: the function and operation overlap, but the trigger and violated invariant differ.
The audit does not validate neighboring findings.
In curl CVE-2026-3805 (\S~\ref{sec:results}), GLM's report notes that the request path points into connection-owned memory but concludes that this is a leak and a latent hazard, not a live use-after-free.
Sonnet~4.6 credits the report, whereas Opus~5.5 rejects it, consistent with our adjudication.

\paragraph{Why target-alarm measurement needs two questions.}
Let $M(P,G)$ mean that findings $P$ assert the failure in reference $G$, and let $V(S,G)$ mean that snapshot $S$ supports that failure; subscript $f$ marks a fixed companion.
Crediting a buggy-task detection requires both: $M(P,G)\land V(S,G)$.
On a correctly fixed companion, $V(S_f,G)$ is false.
The conjunction is therefore false even when the agent incorrectly asserts the removed bug.
A target-alarm test instead asks $M(P_f,G)$ after independently establishing that the fix removed the target.
The task-level verdict does not expose $M(P_f,G)$, so it gives no target-alarm rate or pair accuracy.

% Appendix presentation rows: scripts/build_appendix_tables.py; frozen per-case records remain unchanged.
% Labels are prefixed app:results-/tab:results- except tab:class and tab:fixed, which other sections cite.
\FloatBarrier
\section{Additional Results}
\label{app:results-more}

These results use the frozen 265-case roster and \judgeModel{} verdicts.
A positive saved verdict denotes detection.
Missing or incomplete runs, errors, and empty reports count as misses.
All eight direct-review agents and Codex Security completed every pair; Anthropic's Defending Code harness completed 260 of the 265 pairs.
We release per-case records and analysis code.

\subsection{Run Accounting and Configuration}
\label{app:results-accounting}

\looseness=-1
Every final report has a verdict; empty reports receive a negative verdict without a judge call.

Empty structured reports are not distinguished by cause; only filed findings are evaluated.
Usage comes from each run's session artifacts; output tokens are comparable only within a backend.
Dollar figures are nominal estimates and exclude judging, task construction, and human adjudication.

\paragraph{Agent configuration.}
\label{app:agent-config}

The five Claude Code agents run headlessly with built-in tools and our reporting tool; the three GPT agents use the Codex CLI's non-interactive mode.
Opus uses \code{high} reasoning effort, the other Claude Code agents use their default setting, and the GPT agents use \code{xhigh}.
The non-Anthropic models in Claude Code are served through Anthropic-compatible endpoints.
Claude Code 2.1.283 is used for the Opus~5.5 reviewer, judge, and Anthropic harness; Codex-based runs use releases in the 0.149--0.160 series.
The released artifacts record exact versions and container image pins per arm.
Runs have no turn or token budget and time out after 60 minutes.
Each run reviews an isolated snapshot with Git history removed, using file-reading, search, and shell tools, with built-in web tools disabled.

\subsection{Paired Comparisons}
\label{app:pairs}

We compare direct-review agents on all 265 tasks using exact two-sided McNemar tests, counting missing or errored runs as misses.
No pair among the top four agents (Grok, GPT-6 sol, GLM, and Opus) differs significantly even before correction for multiple comparisons.
After Holm adjustment across all 28 direct-review pairs, no difference is significant within the top four or among Sonnet, GPT-5.6 sol, and GPT-6 luna; all remaining pairs have $p<0.001$.
Collapsing tasks that share an introducing commit into 254 clusters, each credited when any (or every) member is detected, leaves Grok's comparisons with the other top agents nonsignificant even without adjustment.
The other splits in this appendix use exploratory sign, Fisher, Mann--Whitney, or Spearman tests without multiple-comparison correction.

To see where agents separate, we group targets by how many of the other seven direct-review agents find them.
On easy targets, found by six or seven others, every agent except MiniMax~M3 finds at least 90\%.
On contested targets, found by one to three others, the top four find 48--57\%, while Sonnet, GPT-5.6 sol, and GPT-6 luna find 10--19\% and MiniMax~M3 finds 5\%.

\subsection{Results by Task Characteristics}
\label{app:splits}

\paragraph{Project.}
\looseness=-1
V8 has the lowest average direct-review recall (45.2\%) and nine of the 23 unfound targets.
Wasmtime, FreeType, and Redis each have fewer than 20 cases, limiting project-level conclusions.

\paragraph{Bug class.}
Table~\ref{tab:class} groups the reference bug types into memory safety (use-after-free, double free, out-of-bounds access, buffer overflow, null dereference, type confusion, uninitialized memory, wild pointer), logic error, integer overflow, and concurrency (data race or race condition).
Averaged over the eight direct-review agents, recall is 51.5\% on logic errors and 73.1\% on memory-safety flaws.
Logic errors account for 13 of the 23 unfound targets.
Class and project are entangled: about a third of the logic errors are in V8.
Outside V8, logic-error recall remains lower than memory-safety recall for every agent, by 4--22 percentage points, although several differences are not statistically clear.
The concurrency group is too small for class-level conclusions.
With the same harness and reasoning effort, GPT-6 sol finds 58 targets that GPT-5.6 sol misses and misses four that it finds.
GPT-6 sol's largest class gains are 34.1 points on integer overflows and 24.4 points on logic errors.

\begin{table*}[t]
\centering\footnotesize
\setlength{\tabcolsep}{3pt}
\begin{tabular}{@{}lrrrrrrrrrr@{}}
\toprule
Class & $n$ & Opus & Sonnet & Grok & GLM & MiniMax & GPT-6s & GPT-5.6 & GPT-6l & Codex \\
\midrule
% Generated by scripts/build_appendix_tables.py from the frozen 265-case per-case records.
Memory safety & 125 & 85.6 & 68.8 & 84.0 & 85.6 & 40.0 & 82.4 & 69.6 & 68.8 & 60.8 \\
Logic error & 86 & 61.6 & 44.2 & 69.8 & 62.8 & 18.6 & 67.4 & 43.0 & 44.2 & 33.7 \\
Integer overflow & 44 & 68.2 & 63.6 & 84.1 & 70.5 & 45.5 & 79.5 & 45.5 & 45.5 & 43.2 \\
Concurrency & 10 & 60.0 & 50.0 & 50.0 & 60.0 & 0.0 & 60.0 & 40.0 & 30.0 & 20.0 \\
\bottomrule
\end{tabular}
\caption{Recall (\%) by reference bug class; $n$ is the row denominator.
Abbreviations: Opus~5.5, Sonnet~5, Grok~4.6, GLM~5.3, MiniMax~M3, GPT-6 sol (GPT-6s), GPT-5.6 sol, GPT-6 luna (GPT-6l), and Codex Security.}
\label{tab:class}
\end{table*}

\paragraph{Change size.}
The median buggy change touches \medianFiles{} files and \medianLines{} changed lines; 106 of the \numBuggy{} buggy changes touch more than five files.
Every agent has lower recall on changes touching more than five files, although the difference is not statistically clear for GPT-5.6 sol or Codex Security.
The number of agents that find a target has a weak negative association with changed lines and files, which also vary by project.

\paragraph{Vulnerability age.}
\looseness=-1
The introduction-to-fix interval has no clear marginal association with detection for any agent.
This analysis does not control for project or bug class, both of which vary with age; the within-V8 comparison is also inconclusive.

\subsection{Detection Failures}
\label{app:miss-anatomy}

Table~\ref{tab:results-missanatomy} classifies each scanned miss from its trajectory and report.
A task's \emph{root-cause files} comprise the reference finding's location, affected files, and additional files annotated as necessary to explain the mechanism.
For three tasks whose annotated paths do not match changed files at the introducing commit, we use the diff files instead.
A file is \emph{opened} when a read, search, or shell command names it; Codex Security is classified from its shell log.
Each miss takes the first matching category: (A) never opened a root-cause file; (B) opened one but filed no finding; (C) filed findings only elsewhere; or (D) filed a finding in a root-cause file but the judge rejected its mechanism.
Because the diff exposes root-cause files for nearly all tasks, category A is rare by construction and says little about search behavior.
About 68\% of pooled direct-review misses fall in category D.
Recall is also lower when a mechanism spans more root-cause files: pooled over the eight agents, it is 69.6\% for targets with one root-cause file and 38.5\% for targets with five or more.
The drop persists within diff size: from 76.5\% to 55.1\% (one vs.\ two or more root-cause files) among diffs of two to five files, and from 59.0\% to 38.6\% (one vs.\ five or more) among diffs of six or more files.
Outside V8 the pattern is similar (70.3\% to 51.6\%), but only eight such tasks span five or more root-cause files.
Root-cause files include annotated supporting files, and this split is observational.

\begin{table*}[t]
\centering\small
\setlength{\tabcolsep}{4pt}
\begin{tabular}{@{}lrrrrr@{}}
\toprule
Agent & Misses & \makecell[b]{(A) Never\\opened} & \makecell[b]{(B) No\\finding} & \makecell[b]{(C) Only\\elsewhere} & \makecell[b]{(D) Same file,\\other mechanism} \\
\midrule
% Generated by scripts/build_appendix_tables.py from the frozen 265-case per-case records.
Claude Opus~5.5 & 69 & 5.8 & 7.2 & 2.9 & 84.1 \\
Claude Sonnet~5 & 108 & 5.6 & 7.4 & 11.1 & 75.9 \\
Grok~4.6 & 58 & 0.0 & 8.6 & 12.1 & 79.3 \\
GLM~5.3 & 67 & 0.0 & 1.5 & 6.0 & 92.5 \\
MiniMax~M3 & 179 & 1.1 & 29.1 & 8.4 & 61.5 \\
GPT-6 sol & 63 & 0.0 & 22.2 & 15.9 & 61.9 \\
GPT-5.6 sol & 117 & 0.0 & 35.0 & 11.1 & 53.8 \\
GPT-6 luna & 118 & 0.0 & 35.6 & 8.5 & 55.9 \\
Codex Security & 139 & 0.0 & 51.8 & 9.4 & 38.8 \\
\midrule
Eight direct-review agents & 779 & 1.5 & 21.6 & 9.4 & 67.5 \\
\bottomrule
\end{tabular}
\caption{Scanned misses by category (\% of each row's misses).
Categories A--D are mutually exclusive; totals may differ slightly from 100\% due to rounding.
The pooled row includes the eight direct-review agents.}
\label{tab:results-missanatomy}
\end{table*}

In one category-D miss, Opus~5.5's report on SQLite CVE-2025-29088 overlaps the target's allocation logic, but the judge rejects it for describing pointer underflow in a different loop rather than the target mechanism.
Category D rests on judge verdicts we did not re-adjudicate; most such reports offer competing diagnoses rather than explicitly claiming the target site is safe.
About half name a root-cause function, although function names are taken at fix time and can differ at introduction.
Trajectories do not always record subagent reads, so category A is an upper bound.

Median output-token use is higher on missed tasks for every direct-review agent, but also rises with diff size and, for Opus and Grok, with the number of other agents that miss the task.
Within difficulty strata, the Opus difference is no longer statistically clear; Grok's differences vary by stratum and include small miss samples.
These results are observational, not causal.

\paragraph{Unfound targets.}
\label{app:never-found}

\looseness=-1
Grok, GPT-6 sol, and GLM together find 238 targets (89.8\%); adding Opus brings the total to 241, and all eight find 242 (91.3\%).
Codex Security also misses all 23 remaining targets.
Nearly all nevertheless draw findings from at least four direct-review agents, and none draws only empty reports.
Independent detections at each agent's observed recall would yield fewer than one universally missed target in expectation.
The observed excess could reflect task difficulty, hard-to-satisfy reference criteria, shared judge errors, or shared training data; this comparison cannot distinguish these explanations.

\subsection{Report Order}
\label{app:report-order}

\looseness=-1
Among multi-finding direct-review detections with a resolved rank, the credited finding is first in 59.2\% (33.1\% under random ordering) and among the first two in 83.4\%.
Including single-finding reports, reading only the first finding reaches the target in 74.9\% of detections with a resolved rank.
Unresolved ranks are excluded from these percentages; the Hit@1 bounds in Table~\ref{tab:main} count none or all of them as first.
The judge sees findings in the agent's reported order, so a judge position effect cannot be excluded.

\subsection{Fixed Companions}
\label{app:companions}

\paragraph{Volume.}
Mean findings per scan fall for every direct-review agent except MiniMax (Table~\ref{tab:fixed}).
In paired sign tests, Sonnet's decrease and MiniMax's increase are not statistically clear.
Companion and buggy diffs have similar sizes, and median output tokens rise for every agent, so fewer findings do not reflect lower token use.
Pooled over the eight direct-review agents, when a buggy report's only finding is credited, the companion report is empty 44.0\% of the time, against 16.0\% when that finding is rejected.
After a category-D miss (Table~\ref{tab:results-missanatomy}), 88.6\% of companion reports keep a finding in a root-cause file.

\paragraph{Location-only credit.}
Of the nonempty direct-review companion reports, 92.3\% place a finding in a root-cause file.
Crediting any such finding would accept 74.9\% of direct-review companion scans, more than the 63.3\% of buggy scans the judge credits (mean recall in Table~\ref{tab:main}).

\begin{table*}[t]
\centering\small
\setlength{\tabcolsep}{5pt}
\begin{tabular}{@{}lrrr@{}}
\toprule
& \multicolumn{2}{c}{Mean findings per scan} & \\
\cmidrule(lr){2-3}
Agent & Buggy & Fixed & Fixed nonempty (\%) \\
\midrule
% Generated by scripts/build_appendix_tables.py from the frozen 265-case per-case records.
Claude Opus~5.5 & 2.6 & 2.2 & 85.3 \\
Claude Sonnet~5 & 2.2 & 2.1 & 92.8 \\
Grok~4.6 & 3.1 & 3.0 & 92.5 \\
GLM~5.3 & 3.6 & 3.4 & 97.4 \\
MiniMax~M3 & 2.2 & 2.4 & 75.8 \\
GPT-6 sol & 3.1 & 2.7 & 82.6 \\
GPT-5.6 sol & 1.7 & 1.2 & 59.6 \\
GPT-6 luna & 1.6 & 1.2 & 63.4 \\
Codex Security & 1.2 & 0.9 & 54.3 \\
\bottomrule
\end{tabular}
\caption{Report volume across 265 completed pairs per agent.
Fixed nonempty: companions with at least one finding.
No fixed companion receives a positive target verdict.}
\label{tab:fixed}
\end{table*}

\paragraph{What the judge records.}
No direct-review or Codex Security companion verdict is positive under \judgeModel{}.
A positive verdict requires the target mechanism to remain in the reviewed code; a negative verdict does not establish whether a report reasserts the removed target or raises a different issue.
A fixed companion therefore controls only for its designated target; these verdicts do not measure report precision.

\paragraph{Concrete companion examples.}
Two cases from our audit illustrate this distinction (Appendix~\ref{app:judge-audit}).
In OpenSSL CVE-2025-15467, the companion bounds the ASN.1 IV copy and rejects lengths above \code{EVP\_MAX\_IV\_LENGTH}.
GLM instead alleges that a shorter IV leaves bytes uninitialized.
Both judges and our adjudication reject target detection: this allegation differs from the repaired oversized-IV stack overflow.

In curl pull request~\#21652, the companion checks the hexadecimal parser's return value and returns \code{HOST\_NAME} on failure, rejecting the target's empty terminal \code{0x} component.
Grok instead reports ordinary trailing-dot normalization and percent-encoded dots, calling the latter an incomplete fix.
Both judges and our adjudication reject the target match; the audit does not assess whether these neighboring issues are real.

\subsection{Review Harnesses}
\label{app:results-codex}

\paragraph{Codex Security.}

\looseness=-1
Codex Security uses the plugin's diff-scan skill~\citep{openai2026codexSecurityDocs} through the Codex CLI~\citep{openai2026codexcli}, with GPT-5.6 sol at \code{xhigh} effort.
Multi-agent delegation, web search, and sandbox network access are disabled.
Scope rules restrict review to the last commit's diff and directly supporting code and forbid long build or test commands.
Every buggy run completed, with no finding marked as dropped out of scope.
Codex Security files no finding on 21 of the 31 targets found only by direct-review GPT-5.6 sol.
Conversely, it finds nine targets that direct-review GPT-5.6 sol misses (exact McNemar $p<0.001$).
Its mean estimated cost per scan, \$3.74, is 5.1 times that of direct-review GPT-5.6 sol (Table~\ref{tab:main}).
Our benchmark measures recall against one reference per task, but not precision on ordinary changes, which diff-scoped scanners are designed to review.

\paragraph{Defending Code reference harness.}
\label{app:results-anthropic-harness}

Anthropic's Defending Code reference harness~\citep{anthropicDefendingCodeHarness}, using Opus~5.5 at high effort, completed 260 of the 265 pairs.
On these 260 buggy tasks, its recall is 81.5\%, versus 73.8\% for direct-review Opus; it alone detects 31 targets, and direct review alone detects 11 (exact McNemar $p=0.003$).
It files more findings at higher cost (Table~\ref{tab:results-harness}).
Of the 23 targets that no direct-review agent finds, it completed 22 and detects six, three of them concurrency flaws; its credited reports on these six carry 14--32 findings each.
On the 260 completed tasks, its Hit@1 upper bound is 54.6\%, the same as direct-review Opus's.
Counting the five missing pairs as misses gives 80.0\% recall over all assigned tasks.
Because the two arms differ in prompts and budgets, this comparison is descriptive.
On companions, the harness averages 6.2 findings per scan and receives no positive target verdict; these verdicts do not establish report precision.

\begin{table*}[t]
\centering\small
\setlength{\tabcolsep}{5pt}
\begin{tabular}{@{}lrrrr@{}}
\toprule
Arm & Recall (\%) & Mean findings & Median \$ & Median min. \\
\midrule
% Generated by scripts/build_appendix_tables.py from the frozen 265-case per-case records.
Anthropic harness, Opus~5.5 & 81.5 & 8.4 & 3.65 & 7.6 \\
Direct Opus~5.5, same tasks & 73.8 & 2.5 & 0.35 & 1.1 \\
\bottomrule
\end{tabular}
\caption{Anthropic's reference harness and direct-review Opus~5.5 on the same 260 completed buggy tasks.
Findings, nominal cost, and time are per scan; five assigned pairs are missing.}
\label{tab:results-harness}
\end{table*}

\subsection{Judge Output Fields}
\label{app:judge-fields}

\looseness=-1
All saved \judgeModel{} verdicts for the direct-review agents and Codex Security are complete and match the frozen task and report hashes.
Empty reports receive negative verdicts without a judge model call.
The judge receives the task type (buggy task or fixed companion).
Its self-reported confidence is never low, and its auxiliary all-criteria-met flag almost always matches the verdict.
Although credited findings usually receive high match scores, some have low scores; we count detections by verdict, not a score threshold.
On identical reports, Opus~5.5 and Sonnet~4.6 differ on 67 buggy verdicts (35 gains and 32 losses with Opus) and one companion verdict (positive only under Sonnet~4.6).
Opus~5.5's own reports change on seven buggy tasks (four gains and three losses), as do Sonnet~5's, while GPT-6 sol and GPT-6 luna gain the most (net eight and six).
Excluding deterministic empty-report negatives, agreement is \judgeStabilityAgreement{} across \numJudgeStabilityReports{} nonempty buggy reports from the eight direct-review agents and Codex Security.
This sensitivity comparison excludes the Anthropic harness because the two judge evaluations cover different task populations.

\subsection{Temporal Comparisons}
\label{app:temporal}

\begin{figure*}[t]
\centering
\includegraphics[width=\textwidth]{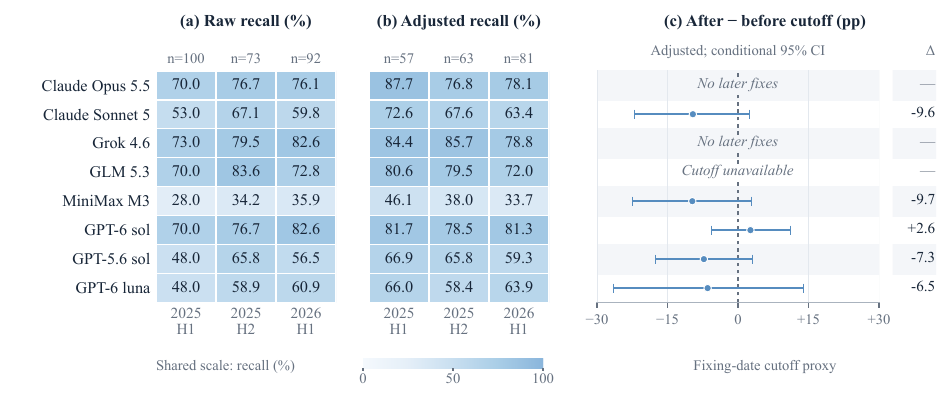}
\caption{Direct-review recall by time: (a)~all cases, (b)~shared project--bug-class groups with fixed weights, and (c)~within-model cutoff differences with conditional 95\% cluster-bootstrap intervals.
Headers give sample sizes.}
\label{fig:temporal}
\begin{minipage}{\textwidth}
\centering\footnotesize
\setlength{\tabcolsep}{3pt}
\begin{tabular}{@{}llrrrr>{$}r<{$}>{$}r<{$}@{}}
\toprule
& & \multicolumn{2}{c}{Cases: before/after} & \multicolumn{2}{c}{Adjusted recall (\%)} & \multicolumn{2}{c}{After $-$ before (pp)} \\
\cmidrule(lr){3-4}\cmidrule(lr){5-6}\cmidrule(l){7-8}
Model & Cutoff & All & Retained & Before & After & \multicolumn{1}{r}{$\Delta$} & \multicolumn{1}{r}{95\% CI} \\
\midrule
% Generated by scripts/build_temporal_analysis.py from the frozen 265-case snapshot.
Claude Sonnet 5 & Jan. 2026 & 173/82 & 121/75 & 71.3 & 61.7 & -9.6 & [-22.0, +2.4] \\
MiniMax M3 & Jan. 2026 & 173/82 & 121/75 & 42.7 & 33.0 & -9.7 & [-22.4, +2.9] \\
GPT-6 sol & Apr.~20, 2026 & 222/42 & 136/41 & 79.9 & 82.6 & +2.6 & [-5.7, +11.1] \\
GPT-5.6 sol & Feb.~16, 2026 & 189/76 & 136/74 & 66.0 & 58.7 & -7.3 & [-17.6, +3.1] \\
GPT-6 luna & May~18, 2026 & 235/28 & 104/28 & 63.6 & 57.1 & -6.5 & [-26.5, +13.9] \\
\bottomrule
\end{tabular}
\captionof{table}{Within-model cutoff comparisons.
Counts exclude the cutoff day or month; retained cases share project--bug-class groups across periods.
Recall uses fixed pooled weights.}
\label{tab:temporal-cutoffs}
\end{minipage}
\end{figure*}

Raw recall gains on newer fixes are sensitive to case mix (Figure~\ref{fig:temporal}).
Adjusted comparisons retain project--bug-class groups present in every period and weight their recalls by pooled case shares, fixed across periods; the half-year comparison retains 201 cases.
Bootstrap intervals resample introducing-commit clusters; sparse groups limit precision, and some trend directions change when retained groups need two or more cases per period.

We compare fixing dates against declared knowledge cutoffs for the GPT models and MiniMax~\citep{openai2026gpt56solCutoff,openai2026gpt6solCutoff,openai2026gpt6lunaCutoff,minimax2026m3Cutoff} and against Sonnet's training-data cutoff~\citep{anthropic2026sonnetCutoff}.
Opus and Grok include training data through June 2026, leaving no later fixes~\citep{anthropic2026opusCutoff,xai2026grok46}; GLM has no documented cutoff.
All five within-model differences are inconclusive, with intervals that include zero (Table~\ref{tab:temporal-cutoffs}).
Fixing dates do not establish training exposure or absence of contamination.

\FloatBarrier

% Keep the artifact section together in the next column.
\newpage
\section{Artifacts and Reproducibility}
\label{app:reproducibility}
\label{app:artifact}
\ifarxiv
Code and data: \url{https://github.com/secprbench/secprbench}; results explorer with saved trajectories: \url{https://secprbench.github.io}.
\else
Artifacts are available at \url{https://anonymous.4open.science/r/secprbench-artifacts-34C1/}.
\fi
The released materials contain task metadata, evaluation results, a validation ledger with source hashes, and code to rebuild the tables without model calls.
Scoring and report-volume accounting use the same final reports, matched to task inputs and saved verdicts by hashes.
The release also includes the complete pairwise comparisons, task-level breakdowns, and finding-order records summarized above.
Replaying validation requires the referenced source trees, builds, and PoCs; reconstructing model outputs also requires model access, the original assembled prompts, and harness artifacts.
Rebuilding the tables does not re-check the judge's verdicts.
\end{document}